# Twitter Subjective Well-Being Indicator During COVID-19 Pandemic: A Cross-Country Comparative Study.

Tiziana Carpi[*] Airo Hino[†] Stefano Maria Iacus[‡§] Giuseppe Porro[¶]

**Abstract**

This study analyzes the impact of the COVID-19 pandemic on the subjective well-being as measured through Twitter data indicators for Japan and Italy. It turns out that, overall, the subjective well-being dropped by 11.7% for Italy and 8.3% for Japan in the first nine months of 2020 compared to the last two months of 2019 and even more compared to the historical mean of the indexes. Through a data science approach we try to identify the possible causes of this drop down by considering several explanatory variables including, climate and air quality data, number of COVID-19 cases and deaths, Facebook Covid and flu symptoms global survey, Google Trends data and coronavirus-related searches, Google mobility data, policy intervention measures, economic variables and their Google Trends proxies, as well as health and stress proxy variables based on big data. We show that a simple static regression model is not able to capture the complexity of well-being and therefore we propose a dynamic elastic net approach to show how different group of factors may impact the well-being in different periods, even over a short time length, and showing further country-specific aspects. Finally, a structural equation modeling analysis tries to address the causal relationships among the COVID-19 factors and subjective well-being showing that, overall, prolonged mobility restrictions, flu and Covid-like symptoms, economic uncertainty, social distancing and news about the pandemic have negative effects on the subjective well-being.

## 1 Introduction

The aftermath of the worldwide Covid-19 outbreak is involving all the aspects of civil coexistence and modifying - we do not yet whether temporarily or permanently - our perception of life. This may result both from a more intense feeling of precariousness, from the restrictions imposed to our social interactions or from objective traces of impoverishment, due to the forced slowdown of economic activities.

Plenty of studies have been carried out, in this short span of time, about the effect of pandemic on feeling, mood and health status - in particular, on mental health - both among Italian (Marazziti et al. (2020), Maugeri et al. (2020), Rossi et al. (2020), Sani et al. (2020), Gualano et al. (2020)) and Japanese people (Yamamoto et al. (2020), Qian and Yahara (2020), Ueda et al. (2020)). Some of these studies, moreover, focus on specific population targets such as elderly and young people or unemployed workers (Orgilés et al. (2020), Shigemura et al. (2020)). Particular attention is devoted to vulnerable categories and people involved in Covid-care activities: e.g., health care workers, who may suffer heavy emotional distress and even discrimination and stigmatization effects (Shigemura and Kurosawa (2020), Asaoka et al. (2020), Torricelli et al. (2020)); pregnant women and newborns (Haruna and Nishi, 2020); patients with specific pathologies (Capuano et al., 2020).

All these studies aim at evaluating the impact of the pandemic on individual and collective well-being and suggesting intervention priorities.

On the other hand, all the economic forecasts agree on the heavy consequences the pandemic is going to have - not only in the short run - on global GDP, consumption, employment and stock market values, even if quantitative estimates of the impact and its distribution over time are still quite uncertain[1] (Chudik et al., 2020; Baldwin and Weder di Mauro, 2020). The studies emphasize similarities and, above all, differences of a global pandemic, compared to previous infectious diseases (Spanish Flu, Ebola virus, Sars, HIV/Aids) and try to estimate the potential economic losses caused by Covid-19 under different scenarios. In fact, when the simulations are carried out, researchers can only rely on data from the first infection wave and the content and duration of restriction measures are still largely unpredictable (Deb et al., 2020): that is why, in some cases, the GDP growth in 2020 is expected between -4.5% and -12.9% in Italy and between -3.1% and -9.7% in Japan, to be compared to a median GDP loss of the 30 major economies between 2.8% and 10.7% (Fernandes, 2020); while other studies predict GDP losses in 2020 between 6 and 214 billion dollars in Italy and between 17 and 549 billion dollars in Japan (McKibbin and Fernando, 2020).

[*]Department of Studies in Language Mediation and Intercultural Communication University of Milan, Milan, Italy.
[†]School of Political Science and Economics, Waseda University, Tokyo, Japan
[‡]Corresponding author, Email: stefano.iacus@ec.europa.eu
[§]European Commission, Joint Research Centre, Via Enrico Fermi 2749, 21027 Ispra (VA), Italy
[¶]Dipartimento di Diritto, Economia e Culture, Università degli Studi dell'Insubria, Como, Italy.

[1]While we write these notes, the second wave of the virus spread is still ongoing and its aftermath will depend, among other things, on the efficacy of the anti-crisis measures.

Despite the evidence that all the sectors are involved in the economic slowdown, the analyses agree on expecting heavier losses in the service sector (travel, tourism and hospitality related activities) and in international trade (Nicola et al., 2020).

This study is a cross-country comparative analysis of the reaction - in terms of collective well-being - of Italy and Japan to the Covid-19 outbreak. In order to examine the evolution of well-being, we adopt a subjective well-being indicator based on Twitter data for Italy (Iacus et al., 2019, 2020a,b) and Japan (Carpi et al., 2020). The study finds a perceived well-being dropdown of 11.7% in Italy and 8.3% in Japan during the first nine months of 2020. Further analysis focuses on the determinants of what impacted the two indicators through the Covid-19 pandemic dynamically: this allows to show that the relative importance of explanatory variables in determining subjective well-being changes in a very short time, as the socio-economic and public health conditions show high instability. The two countries show differences both in their exposition to the pandemic and in their language and cultural background: the comparison will also contribute to disentangling these two aspects in the evolution of subjective well-being. A structural equation model approach shows that, overall, prolonged mobility restrictions, flu and Covid-like symptoms, economic uncertainty, social distancing and news about the pandemic have negative effects on the subjective well-being.

The work is organized as follows: Section 2 is a review of the different approach to big data analysis in the field of subjective well-being studies and Section 3 introduces the indexes used in this study. Section 4 describes the time series of the two subjective well-being indexes considered in the analysis during the Covid-19 pandemic. Section 5 introduces the set of additional covariates that will be used as control variables in Section 6, where the dynamic analysis of the impact of each variable is analysed. Section 7 attempts to study the complexity of subjective well-being through an application of structural equation modeling building upon the evidence of the previous sections. Section 8 draw conclusions based on the results coming from the previous sections. Finally, Section 9 describes the limit of this work and presents further possible developments.

# 2 Big Data, Well-Being and Covid-19

Large-scale dataset provided by social networking sites (SNS) offer an unprecedented information source for monitoring public opinion and, particularly, self-perceived quality of life and well-being on an individual and - more interestingly - collective base. Compared to traditional sources, data originated by SNS offer undoubted advantages: they are available in real time, continuously updated and less expensive to collect: that is the reason why they are often used for studying real-time changes in emotional mood as a reaction to unexpected and extraordinary occurrences (Miura et al., 2015; Jones et al., 2016), like the Covid-19 pandemic. On the other hand, these data also show objective limitations: above all, they come from a community, which - despite its large dimension - is not necessarily representative of the whole population of a country (Grewenig et al., 2018; Kern et al., 2016).

Twitter is one of the most popular SNS, with 330 million monthly active users worldwide in 2019[2]. Due to the brevity of the messages allowed, Twitter is often assumed to be one of the most suitable platform to evaluate emotional well-being. Recent literature provides some examples of well-being evaluations that rely on Twitter data and, based on sentiment analysis methods, aim at monitoring the day-by-day evolution of self-declared emotional status of a community.

Dodds et al. (2011) built an happiness indicator, called *hedonometer*, based on a so-called 'closed vocabulary' approach: they measured the frequency of use of a set of ten thousand words for which they have obtained happiness evaluations on a nine-point scale, using Mechanical Turk[3]. Their dataset was huge, being made of around 4.6 billion expressions posted by over 63 million Twitter users from September 2008 to September 2011. The project is still ongoing and the hedonometer is now evaluated daily by the University of Vermont Complex Systems Center, which can provide, therefore, a time series since 2008[4] mainly for text written in English. It is worth noting that, through a graphical inspection, it appears that no particular trend is observed in 2020 for this index meaning probably that the *hedonometer* index does not capture what we aim at in this study.

A subjective well-being indicator - named *Gross National Happiness index* - has been proposed by Rossouw and Greyling (2020): the indicator is evaluated since 2019 in three Commonwealth member countries: South Africa, New Zealand and Australia[5]. The aim of the project is to measure, in real time, the sentiment of countries' citizens during different economic, social and political events: its first application has been an examination of the well-being impact of social restrictions imposed during the first wave of Covid-19 pandemic in South Africa (Greyling et al., 2020). In order to calculate the index, the sentiment analysis is applied to a live Twitter-feed, and each tweet is assigned either a positive, neutral or negative sentiment. Then, an algorithm evaluates a happiness score on a 0-to-10 scale. The Gross National Happiness index provides an happiness score per hour for each of the three countries. The approach adopted by the Gross National Happiness index is, in many respects, similar to the methodology followed by the index proposed and applied in this paper. In both cases, in fact, the sentiment is drawn from a reading and classification of tweets, that involves the work of human coders, while the evaluation of collective well-being is entrusted to an algorithm. Nevertheless, the evaluation method of the Gross National Happiness index still relies, to some extent, on the meaning of the words used in the texts to express a feeling, particularly when the analysis goes deep into the breakdown of sentiment into specific positive or negative feelings. On the contrary, the methodology and the algorithm we propose here make use of all the terms (not only the words with a particular meaning) and the language structure to infer mood, hence exploiting suggestions coming also from jargon, neologisms, paradoxical or ironic expressions (Ceron et al., 2016).

[2]https://statista.com
[3]https://www.mturk.com/
[4]https://hedonometer.org/timeseries/en_all/
[5]https://gnh.today

Although not based on Twitter data, a particular mention is due to Algan et al. (2016), who propose an indicator of subjective well-being in the United States based on Google Trends data from January 2008 to January 2014. The authors retrieve from Google Trends several groups of keywords whose behavior over time fits with the weekly time series of well-being measures based on surveys provided by Gallup Analytics[6]. Moreover, a model is built to predict - via Google Trends - the well-being values obtained through surveys: the good out-of-sample performance of the model allows for suggesting the use of web-based indicators to complement survey-based or official statistics, exploiting the advantages of web data in terms of high frequency and geo-localization. The study also identifies the keyword categories (job search, financial security, family life, leisure) that better predict well-being variations. It is worth noting that this indicator has much in common with the subjective well-being index we propose in this paper: in fact, both the indicators aim at improving official statistics, providing high frequency and local estimates of well-being, usually not available at an official level. This notwithstanding, they differ in that Algan et al. (2016)'s indicator reliability is twice dependent on external sources: on one hand, it adopts a sort of 'closed vocabulary' approach, based on the goodness-of-fit of Google Trends with Gallup Analytics surveys; on the other hand, Gallup Analytics is also the term of comparison to test the forecasting ability of the indicator. On the contrary, our index - as a sentiment analysis instrument - tries to extract the feeling directly from the language used by the web surfers.

Also not based on Twitter, but strictly related to the topic of this research is a survey measure of subjective well-being which, along with Google Trends data, underpins the work by Foa et al. (2020), where the authors examine the evolution of well-being in Great Britain during the first wave of the pandemic, in spring 2020. The original contribution of this study is in his attempt to disentangle the well-being effect of the pandemic from the effect of the lockdown restrictions, showing that, while the Covid-19 outbreak negatively affects self-perceived well being, the imposition of lockdown measures has a positive impact on well-being, likely due to a better work-life balance, an increase in remote worker autonomy, the adoption of government support schemes. Google Trends series allow for an extension of the results to countries other than Great Britain.

The significant shift in households' economic sentiment in the 27 EU countries, following the Covid-19 outbreak and ensuing lockdown restrictions, is investigated by van der Wielen and Barrios (ming). The research, similarly to the previous ones, relies on panel dataset coming from Google Trends, covering business cycle, labour market and consumption related queries. As a result, the study reveals a substantial worsening of the sentiment along all the investigated dimensions in the months after the pandemic outbreak. The change of the sentiment is more pronounced in the EU countries where the economic and labor market conditions were less favourable at the onset of the health crisis, arguably reflecting the fear of a persistent high unemployment level in the post-pandemic phase. Finally, in a comparison with the consequences of the 2008 economic recession, the study shows that the fall in economic sentiment is even more marked during the 2020 pandemic, especially for unemployment-related sentiment, confirming widespread concerns about pervasive and long-lasting consequences of the crisis.

# 3 The Twitter Subjective Well-Being Indicators

In this study we use the subjective well-being indicators introduced in Iacus et al. (2019, 2020a,b) for the Italian case, and in Carpi et al. (2020) for the Japanese case. These indexes are based on semantic extraction of expressions of well-being from Twitter data and are inspired by the Happy Planet Index proposed by the New Economic Foundation think-tank (NEF, 2012). The subjective well-being indicator, respectively called **SWB-I** and **SWB-J** for Italy and Japan, consists of eight dimensions, concerning three different areas of well-being: *personal* well-being, *social* well-being and well-being at *work* as listed in Table 1.

Each component of the SWB index is extracted via human supervised sentiment analysis and, in particular, through the iSA algorithm (Ceron et al., 2016). In this approach the training set is prepared by human coders as no dictionary or semantic rule is applied. This step of the process is crucial in that, being a qualitative classification, all the relevant information is distilled at this stage but it affects directly the extension of the analysis to the whole set of tweets. Each tweet in the training set is classified into four categories `positive`, `neutral`, `negative` and `OffTopic` with respect to each component of the SWB index. Table 2 is an example of coding rules for the `emo`

[6] https://https://www.gallup.com/analytics/home.aspx

| Area | SWB dimension | Description |
|---|---|---|
| Personal well-being | `emo`-tional well-being | the overall balance between the frequency of experiencing positive and negative emotions, with higher scores showing that positive feelings are felt more often than negative ones |
| | `sat`-isfying life | having a positive assessment of one's life overall |
| | `vit`-ality | having energy, feeling well-rested and healthy while also being physically active |
| | `res`-ilience and self-esteem | a measure of individual psychological resources, of optimism and of the ability to deal with life stress |
| | positive `fun`-ctioning | feeling free to choose and having the opportunity to do it; being able to make use of personal skills while feeling absorbed and gratified in daily activities |
| Social well-being | `tru`-st and belonging | trusting other people, feeling treated fairly and respectfully while experiencing sentiments of belonging |
| | `rel`-ationships | the degree and quality of interactions in close relationships with family, friends and others who provide support |
| Well-being at work | quality of `wor`-k | feeling satisfied with a job, experiencing satisfaction with work-life balance, evaluating the emotional experiences of work and at work conditions |

**Table 1:** The eight components of the SWB index.

| Example (En) | Example (JP) | Classification |
|---|---|---|
| how lucky I am ! | ラッキだ！ | positive |
| what a beautiful day :) | 美しく晴れ渡った日 | positive |
| finally I passed the exam! | やっと合格した。 | positive |
| there are good and bad people | いい人と悪い人がいる。 | neutral |
| tonight I have a date with my girlfriend <3 | 今晩彼女とデートする予定 <3。 | positive |
| my girlfriend quit me last night | 昨晩彼女に振られちゃった。 | negative |
| I feel sick and I have headache | 風邪を引いて、頭が痛いんだ。 | negative |

**Table 2:** Example of classification rule from fictitious texts with the aim of classifying the *emotional* (`emo`) component of the *personal* well-being.

component on fictitious tweets, while the following is an example of actual tweet classification (original Japanese version on the left, approximate English translation on the right):

| | |
|---|---|
| "体中が痛い……<br>下手くそな証拠だ……<br>いくちゃんからメール来てたから元気でた<br>よし、行動しよ" | *"My body hurts...<br>Bad shit...<br>I was fine because I received an email from Iku-chan.<br>OK, act"* |

it has been categorized under *resilience and self-esteem* (`res`) as *positive*, as well as this one:

| | |
|---|---|
| "意識高い高いと自尊心高い高いしてみたい" | "High consciousness and high self-esteem" |

More complex are tweets like this one:

| | |
|---|---|
| "精神とは控えめに見ても 90 パーセント妄想であって、妄想を自己と切り離す作業を日夜続けることが、初期における人間の精神生活の主なノルマである。" | "The spirit is 90% delusion, even if it is conservative, and continuing the work of separating the delusion from the self, day and night, is the main norm of human mental life in the early days." |

it may seem to express a negative view about life and therefore it can be arguably classified as *negative* for the component *satisfying life* (`sat`). And so forth.

The `OffTopic` category comprises all the texts which fall outside of the scope of the analysis and it is an essential input to the machine learning algorithm to disentangle *noise* from *signal*.

Once the training set has been completely hand-coded, the iSA algorithm (Ceron et al., 2016) is applied to daily test sets of data. Each estimated distribution will contain the same four categories mentioned in the above and an index for each component is calculated for day $d$ and for any dimension - e.g. for the `emo` component - as follows:

$$\texttt{emo}_d := \frac{\%\texttt{positive}}{\%\texttt{positive} + \%\texttt{negative}} \in [0, 1]$$

The final $\text{SWB}_d$ index is just the average of the eight components `emo`, `sat`, `vit`, `res`, `fun`, `tru`, `rel` and `wor`. Further technical details can be found in the original references.

# 4 The Decline of SWB-I & SWB-J During COVID-19

The data used in this work come from two different repositories that were collected under two different projects but in both cases using Twitter search API. The Japanese tweets were collected using only the filter on language = `Japanese` and country = `Japan` and similarly for Italy (`Italian` and `Italy`). As per the official documentation,

Twitter search API only provides a 10% sample of all tweets though the company does not disclose any information about the representativeness of the sample with respect to the whole universe of tweets posted on the social network. Nevertheless, according to our personal experience, also confirmed in large scale experiments by Hino and Fahey (2019), the coverage of topics and keywords is quite accurate and appears to be randomly selected: therefore we consider the Twitter data used in this study as a representative sample of what is discussed on Twitter. According to Statista[7] there are about 8 million accounts active daily in Italy whilst about 52 millions are in Japan, therefore the number of tweets posted is not comparable. To keep the volumes of tweets comparable between the two indicators we imposed a maximum number of at most 50,000 tweets per day to our Twitter crawlers. As a result, the total volume of tweets is 13,975,242 for Italy and 12,907,902 for Japan and the data were collected since 2020-11-01 till 2020-10-11 for Italy and 2020-09-20 for Japan. These tweets are part of two separate repositories that were collected for Italy since 2012 and for Japan since 2015. For both projects, systematic download of data was stopped in 2018 and resumed on late 2019 with alternate fortune due to changes in Twitter API limits. For Italy some historical data were collected ex-post for the year 2019 for other research repositories and included in this data collection. We denote by SWB-I the index for Italy (Iacus et al., 2019, 2020a,b) and SWB-J the index for Japan (Carpi et al., 2020) as explained in Section 3. Table 3 shows the yearly average values of the SWB-I and SWB-J indicators since 2012 and 2015 for Italy and Japan respectively. In both cases, the two indicators dramatically dropped (more in Italy than in Japan) in 2020. While the average values can be attributed intuitively to cultural differences in the way positive and negative emotions are expressed in the two countries, clearly the 2020 drop is definitely related to the COVID-19 pandemic, yet with differences between the two countries as it will be explained in the next sections.

| Year | 2012 | 2013 | 2014 | 2015 | 2016 | 2017 | 2018 | 2019 | 2020 | Nov-Dec 2019 vs 2020 |
|---|---|---|---|---|---|---|---|---|---|---|
| SWB-I | 48.9 | 52.2 | 49.7 | 48.7 | 50.5 | 57.7 | 55.7 | 54.1 | 42.4 | -11.7 |
| SWB-J | – | – | – | 54.4 | 53.6 | 53.2 | 52.5 | 35.3 | 27.0 | -8.3 |

**Table 3:** Average values of SWB-I and SWB-J from 2015 till 2020. For Italy, data in 2015 were not available for the whole year.The statistics for 2019 are referred only to the months of November and December. For 2020, the average refers to period from 1st January up to 11th October for Italy and 20th September for Japan.

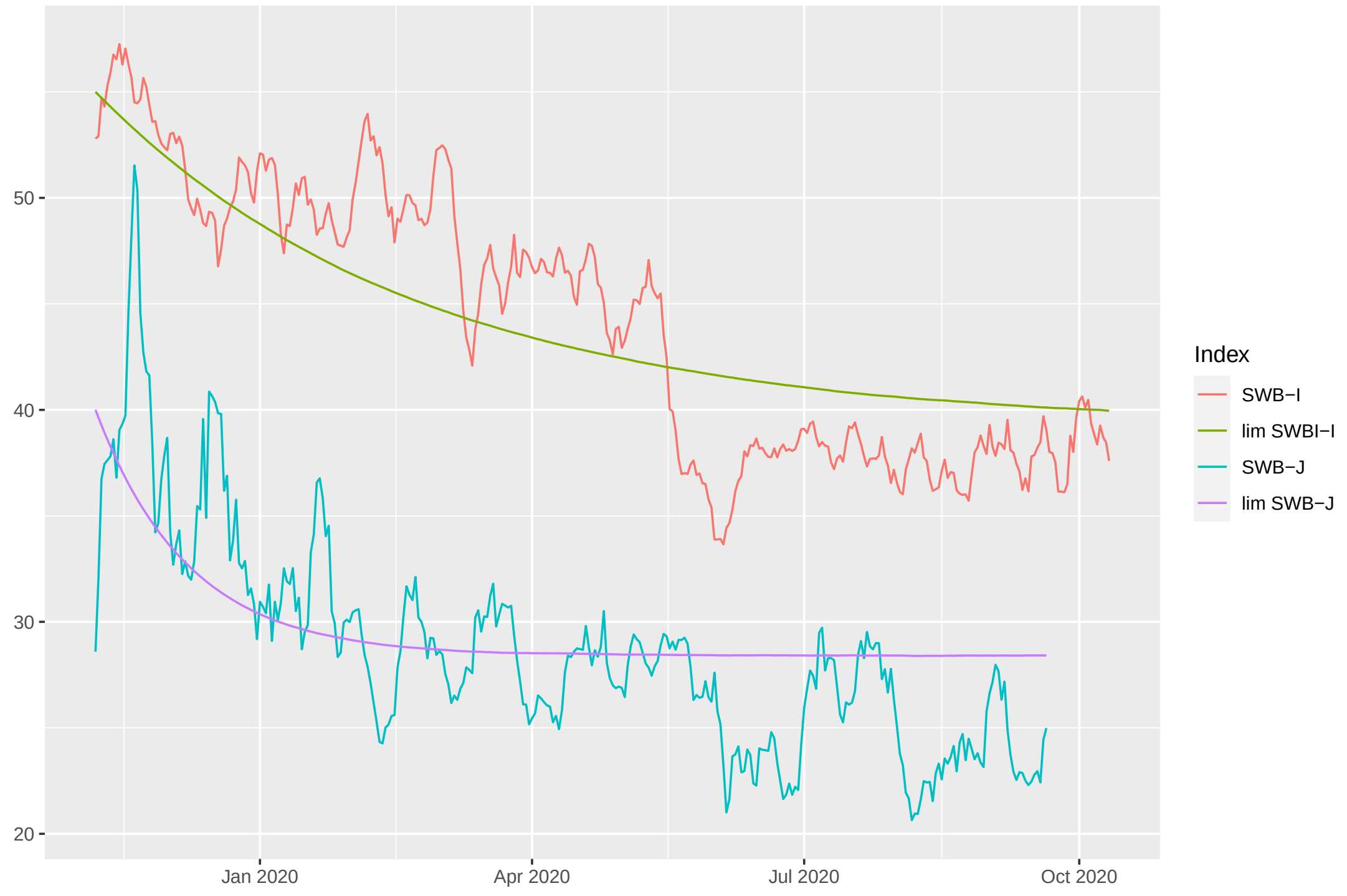

**Figure 1:** SWB-I and SWB-J indexes from November 2019 till 10th October 2020 for Italy and 20th September for Japan with estimated limiting dynamical systems.

In order to compare the two time series in a stochastic modelling approach, we try to fit four different stochastic differential equations models (Iacus, 2008; Iacus and Yoshida, 2018). As the two time series exhibit a negative trend in the first half of 2020 and seem to rebound around a long term mean, we assume a drift coefficient of mean-reverting time. A typical drift form is described as $\alpha(\beta - X_t)$ where $\beta$ represents the long term mean around which the time series $X_t$ oscillates and $\alpha$ is called the *speed* of mean reversion: the higher $\alpha$, the faster the process converges to its long run mean. A first example of this model is the so called Ornstein-Uhlenbeck (Uhlenbeck and Ornstein, 1930) or Vasicek model (Vasicek, 1977) , which is a continuous version of an AR(1) model[8]:

$$\mathrm{d}X_t = \alpha(\beta - X_t)\mathrm{d}t + \sigma \mathrm{d}W_t, \quad X_0 = x_0 \tag{1}$$

where $\mathrm{d}X_t$ is infinitesimal increment of the process between $t$ and $t+\delta$, i.e. $\mathrm{d}X_t = X_{t+\delta} - X_t$, $\mathrm{d}W_t \sim N(0, \mathrm{d}t)$ is the infinitesimal increment of the Wiener process (or Brownian motion) that represents exogenous random Gaussian shocks, $\sigma > 0$ is the scaling factor and $x_0$ is some initial condition. Next is the geometric Brownian motion (Black and Scholes, 1973; Merton, 1973) model:

$$\mathrm{d}X_t = \alpha(\beta - X_t)\mathrm{d}t + \sigma X_t \mathrm{d}W_t, \quad X_0 = x_0 \tag{2}$$

where the term $\sigma X_t \mathrm{d}W_t$ represents the feedback in the system, in the sense that the noise $\mathrm{d}W_t$ interacts with the state of the system $X_t$. We also fit the CIR (Cox-Ingersol-Ross) model (Cox et al., 1985) which differs from the previous model for the $\sqrt{X_t}$ in the diffusion coefficient[9]. The role of the square root is to dump a bit the feedback effect:

$$\mathrm{d}X_t = \alpha(\beta - X_t)\mathrm{d}t + \sigma \sqrt{X_t} \mathrm{d}W_t, \quad X_0 = x_0. \tag{3}$$

[7] https://statista.com

And finally, when one is uncertain about the dumping effect of the feedback term, one can generalize the model to the CKLS (Chan, Karolyi, Longstaff and Sanders) model (Chan et al., 1992) adding an exponent $0 < \gamma < 2$ in the diffusion coefficient:

$$\mathrm{d}X_t = \alpha(\beta - X_t)\mathrm{d}t + \sigma X_t^{\gamma} \mathrm{d}W_t, \quad X_0 = x_0 \tag{4}$$

Clearly, CKLS embeds VAS (for $\gamma = 0$), CIR (for $\gamma = 0.5$) and GBM (for $\gamma = 1$). We estimate these models on the two SWB-I and SWB-J indexes thorough the `yuima` R package (Brouste et al., 2014; Iacus and Yoshida, 2018) via quasi-maximum likelihood estimation. The results of the estimation are given in Table 4. The main characteristics of the first threee models is that VAS model has Gaussian increments, GBM has log-Normal increments and CIR has non-central Chi-Squared increments.

| Index | $\alpha$ | $\beta$ | $\sigma$ | $\gamma$ | Model | AIC |
|---|---|---|---|---|---|---|
| SWB-I | 3.16 | 38.99 | 14.7 | | VAS | 787.0 |
| | (2.41) | (6.22) | (0.56) | | | |
| | 3.57 | 39.54 | 0.33 | | GBM | 782.9 |
| | (2.46) | (4.60) | (0.01) | | | |
| | 3.34 | 39.28 | 2.20 | | CIR | 781.9 |
| | (2.42) | (5.35) | (0.08) | | | |
| | 3.42 | 39.37 | 1.12 | 0.68 | CKLS | 783.6 |
| | (2.44) | (5.08) | $(\cdot)$ | $(\cdot)$ | | |
| SWB-J | 12.92 | 28.44 | 26.25 | | VAS | 1090.0 |
| | (5.52) | (2.19) | (1.05) | | | |
| | 11.46 | 28.43 | 0.83 | | GBM | 1027.4 |
| | (5.71) | (2.23) | (0.03) | | | |
| | 11.98 | 28.42 | 4.62 | | CIR | 1055.1 |
| | (5.61) | (2.21) | (0.19) | | | |
| | 11.64 | 28.41 | 0.05 | 1.84 | CKLS | 1010.7 |
| | (5.93) | (2.22) | (0.01) | (0.08) | | |

**Table 4:** Fitting different SDE models. Standard errors of the estimates are in parentheses and $(\cdot)$ means estimates of variance-covariance matrix did not converge).

Looking at the results in Table 4 it seems clear that the models agree on the $\alpha$ and $\beta$ parameters. Indeed, $\beta$ correspond to about 39 for SWB-I and 28 for SWB-J across models with about 11 percentages points of difference between the two countries, and $\alpha$ varies in the interval [3.16,3.57] for SWB-I and in the interval [11.46,12.92] for SWB-J, meaning that the convergence to the long run mean of the SWB index is between 3 and 4 times faster in Japan than in Italy. Figure 1 show also the estimated limit deterministic dynamical systems obtained taking the limit as $\sigma \to 0$ of the stochastic differential equation models which best fits, in terms of AIC, the set of data (respectively, CIR for SWB-I and CKLS for SWB-J): what these deterministic limit processes show is that the decline of subjective well-being in Italy is much more emphasized, in 2020, compared to what happens in Japan whilst the convergence to the long run mean is faster for Japan.

In summary, the analysis shows that the subjective well-being has a short run (or volatile) and a medium-long run (or structural) component, which are usually intended as emotional well-being *versus* life evaluation. Moreover, the mean is not necessarily stable during the whole year and therefore, a annual single survey and the consequent point estimate are not appropriate to capture the overall level and evolution of well-being of a country.

[8] For $\beta = 0$.

[9] In a stochastic differential equation model the term associated to $\mathrm{d}W_t$ is called *diffusion* coefficient.

# 5 Data Collection of Potential Determinants of the SBW Indexes

Apart from noting a decline of the SWB indexes for both Italy and Japan, this study aims at understanding the impact of several potentially explanatory factors on the SWB indicators for these two countries. For this reason, several data sources have been considered.

## 5.1 Covid-19 Spread Data

On the official statistics side, we obtained from the WHO the data on the number of confirmed Covid-19 cases and deaths. We replaced negative values[10] with zeros and take 7-days moving average to reduce the impact of weekend days late reporting, that induces artificial periodicity in the data.

## 5.2 Financial Data

In order to capture high frequency financial dynamics, we consider the main stock market indexes of both countries, specifically the Nikkei and FTSE MIB index. Daily adjusted closings data are taken from Yahoo! Finance[11] through the `quantmod` package (Ryan and Ulrich, 2020).

## 5.3 Air Quality Data

We then considered air quality data available through the Air Quality Open Data Platform[12] and, in particular, the PM2.5 pollutant concentration and the temperature. Also in this case we considered a 7-days moving average and aggregate data at country level. The two variables roughly capture, on one hand, the amount of pollutant reduction during the pandemic due to the lockdown and, on the other hand, the effect of high/low temperature on mood (Curini et al., 2015).

## 5.4 Google Search Data

Subjective well-being is influenced by micro and macro variables, where the latter are usually captured by official statistics. Both micro and macro observable dimensions are characterized by very low frequency and are available with considerable delay. Therefore, as in Choi and Varian (2012) we adopt as proxies of these variables several types of Google research topics in the two countries available through the Google Trends[13] portal. Data have been downloaded through the `gtrendsR` package[14] (Massicotte and Eddelbuettel, 2020).

Google Trends offers two types of search statistics: one is based on the exact keyword and one is based on the concept of *topic*. The difference is that topics include all search terms related to that topic or, put it another way, topics are a collection of search terms. Topics are normalized across countries, so there is no need to translate a topic into a specific language. We included several research topics, distinguishing between general web searches (blogs, forum, etc) and specific news-related searches. We included terms related to the pandemic, real economy and job market, health conditions and searches for adult content. The latter has been introduced following the suggestions in by Stephens-Davidowitz (2018) that found that Google search for adult content predicts the United States unemployment rate from 2004 into 2011 with the motivation that *"(...) unemployed people presumably have a lot of time on their hands. Many are stuck at home, alone and bored"*. Indeed, *"with the global expansion of the COVID-19 pandemic, social or physical distancing, quarantines, and lockdowns have become more prevalent and concurrently, Pornhub, one of the largest pornography sites, has reported increased pornography use in multiple countries, with global traffic increasing over 11% from late February to March 17, 2020"* (Mestre-Bach et al., 2020). Happyness literature also seem to consider this relationship with, so to say, boring living conditions (D'Orlando, 2011) as well as to COVID-19 specific stress conditions (Döring, 2020).

Table 5 contains the complete list of topics. Topics and keywords are the same for the two countries with two differences: 1) for Japan only we added the search for the keyword Corona 'コロナ' in katakana alphabet, as we noticed a remarkable difference between the topic and this exact search in terms of time series patterns; 2) for Italy only we included the keyword 'Rt' for reproduction number, as this is often reported in the news and in the official statements of the government. The Rt indicator is not associated to the virus in Japan, so we did not include it for Japan.

[10] Negative values exists because the number of cases is obtained through differentiation of the cumulative number of cases, that is updated retrospectively in some cases, leading to negative differences.

[11] https://finance.yahoo.com.

[12] World Air Quality Index (WAQI) https://aqicn.org/data-platform/covid19/.

[13] Google Trends, https://www.google.com/trends.

[14] All the computations have been processed using the R statistical environment (R Core Team, 2020).

| Search term | country | keyword type | framework | type of search |
|---|---|---|---|---|
| Coronavirus | IT/JP | topic | pandemic | web |
| CoronavirusNews | IT/JP | topic | pandemic | news |
| (コロナ) Corona | JP | topic | pandemic | web |
| (コロナ) CoronaNews | JP | topic | pandemic | news |
| Covid | IT/JP | topic | pandemic | web |
| CovidNews | IT/JP | topic | pandemic | news |
| Rt | IT | keyword | pandemic | web |
| Wuhan | IT/JP | topic | pandemic | web |
| Unemployment | IT/JP | topic | economics | web |
| UnemploymentNews | IT/JP | topic | economics | news |
| Economy | IT/JP | topic | economics | web |
| EconomyNews | IT/JP | topic | economics | news |
| GDP | IT/JP | topic | economics | web |
| GDPNews | IT/JP | topic | economics | news |
| Depression | IT/JP | topic | health | web |
| Stress | IT/JP | topic | health | web |
| Insomnia (disorder) | IT/JP | topic | health | web |
| Health | IT/JP | topic | health | web |
| Solitude | IT/JP | topic | health | web |
| AdultContent | IT/JP | topic | leisure | web |

**Table 5:** Google search topics used in the analysis.

## 5.5 Google Mobility Data

In addition to the Google search data, we include the human mobility data obtained through the Google Covid-19 Community Mobility Reports[15]. We considered a 7-days moving average data of the "residential and workplace percent change from baseline" statistic available in the data to capture, roughly, the effect of lockdown restrictions on human mobility.

## 5.6 Facebook Survey Data

Since late April 2020 Facebook, in partnership with several universities, has conducted the "COVID-19 World Survey Data" collecting several indicators. These COVID-19 indicators are derived from global symptom surveys that are placed by Facebook on its platform. The surveys ask respondents how many people in their household are experiencing COVID-like symptoms, among other questions. These surveys are voluntary, and individual survey responses are held by University of Maryland and are shareable with other health researchers under a data use agreement. No individual survey responses are shared back to Facebook. Using this survey response data, the University of Maryland estimated the percentage of people in a given geographic region on a given day who:

- have COVID-like illness (FB.CLI) = fever, along with cough, shortness of breath, or difficulty breathing;
- have influenza-like illness (FB.ILI) = fever, along with cough or sore throat;
- have reported to use mask cover (FB.MC);
- have reported had direct contact (FB.DC), longer than one minute, with people not staying with them in last 24 hours;
- are worried about themselves and their household' s finances in the next month (FB.FH). Only respondents who have replied as being very worried and somewhat worried.

Instead of direct counts that may have missing data for some date, we use the smoothed versions of the indicators based on a seven-day rolling average. Data have been collected through the COVID-19 World Symptom Survey Data API (Fan et al., 2020).

## 5.7 Restriction Measures Data

Finally, we constructed a dummy variable `lockdown` for both countries, taking value 1 when lockdown or other types of restrictions were in force in each country. For Italy[16], a national lockdown was enforced since 9 March 2020 and lifted on 3 June 2020. For Japan[17], there was no strict lockdown, but the state of emergency has been declared starting from 8 April 2020 and lifted on 21 May 2020 for most prefectures. But, as the remaining five prefectures has to wait till 25 May 2020, we decided to set our dummy equal to 1 for Japan for the whole period 2020-04-08/2020-05-25.

Table 6 reports the complete list of variables used in the analysis. It is worth mentioning that, due to the different time coverage of the data, we extended the analysis to 10 October 2020 for Italy and to 20 September 2020 for Japan.

[15] Google LLC "Google Covid-19 Community Mobility Reports", https://www.google.com/covid19/mobility/ Accessed: 2020-10-15.

[16] Source: https://en.wikipedia.org/wiki/COVID-19_pandemic_in_Italy.

[17] Source: https://en.wikipedia.org/wiki/COVID-19_pandemic_in_Japan.

| Variable | area | Source |
|---|---|---|
| SWB-I, SWB-J | subjective well-being | Twitter |
| Solitude | well-being | Google Trends |
| Depression | well-being | Google Trends |
| Stress | well-being | Google Trends |
| Insomnia | well-being | Google Trends |
| Health | health/well-being | Google Trends |
| PM2.5 | health/environment | WAQI |
| Temperature | environment | WAQI |
| Cases | pandemic | WHO |
| Deaths | pandemic | WHO |
| Coronavirus, CoronavirusNews | pandemic | Google Trends |
| (コロナ) Corona, CoronaNews | pandemic | Google Trends |
| Covid, CovidNews | pandemic | Google Trends |
| Rt | pandemic | Google Trends |
| Wuhan | pandemic | Google Trends |
| Unemployment, UnemploymentNews | economy | Google Trends |
| Economy, EconomyNews | economy | Google Trends |
| GDP, GDPNews | economy | Google Trends |
| FTSEMIB | economy | Yahoo! Finance |
| Nikkei | economy | Yahoo! Finance |
| AdultContent | leisure | Google Trends |
| FB.CLI | health/well-being | Facebook |
| FB.ILI | health/well-being | Facebook |
| FB.MC | behavioural | Facebook |
| FB.DC | behavioural | Facebook |
| FB.FH | well-being | Facebook |

**Table 6:** Times series used in the analysis. In the analysis Japanese variable start with *j* and the Italian ones with *i*, e.g., `iCases`, `jCases`.

# 6 What Impacted The Subjective Well-Being Indexes?

Although the SWB-I and SWB-J indicators are composed by a range of different dimensions, it is also true that components like `emo` may be affected by the pandemic in several ways, like, e.g., stress, fear of the virus coming from the news or the statistics of cases and deaths, etc. In this study, we want to investigate and possibly describe the complexity of the determinants of subjective well-being, as measured by the proposed social media indexes, following a data science approach.

## 6.1 Preliminary Correlation Analysis

A simple correlation analysis[18] between each index and the other covariates may lead to counter-intuitive conclusions. Indeed, looking at Figure 2 it is possible to notice that for some variables the correlation with SWB-I (and SWB-J) change monthly, from November 2019 till September 2020, and also with respect to the whole 2020 (last row of each correlation plot in Figure 2). More precisely, looking at Figure 2 top panel for SWB-I, while the variable temperature (`itemperature`) is negatively correlated with well-being in Italy, monthly and for the whole year 2020, news about economy (`iEconomyNews`) have positive correlation up to May 2020, and then it switches sign. Or, again, the variables `iCoronaVirus`, `iCoronaVirusNews`, `iCovid`, `iCovidNews`, `iHealth` are negatively correlated between January and March 2020, then almost all the correlations switch sign. Other variables, like `ilockdown` do matters, with negative correlation, but clearly only in the periods when the lockdown is actually in force. Similar up-and-down trends occur for the Japanese index. It is quite surprising, when we read the panels of Figure 2 by row, noting that groups of variables that show a positive relationship with SWB at some point in time change the correlation sign in a month or so. Clearly, this is the effect of the complexity of the well-being itself and its dynamics over time. More details emerge from the monthly regression analysis.

## 6.2 Monthly Regression Analysis

To take into account the dynamic nature of subjective well-being, shown also by the preliminary correlation study, we run a monthly stepwise regression analysis[19]. Table 8 reports the estimated coefficents and summary analysis for the SWB-I index, whilst Table 9 shows the same analysis for the SWB-J index. In order to save space, the tables show only the months from January to September 2020. The tables also show the results for the whole period January-September 2020.

Considering the Italian case first, it can be seen that the following variables have a negative effect on the yearly scale: `iRt`, `iGDP`, `iHealth`, `iInsomnia`, `iWuhan`, `ilockdown`, `iFB.CLI` and `iFB.DC` have a negative impact on SWB-I, which is quite admissible, but `iCoronavirusNews`, `iCovid`, `iUnemployment`, `iGDPNews`, `iResidential`, `iWorkplace`, `itemperature`, `iFB.ILI` have a positive impact on SBW-I and this is probably the outcome of an average effect, though we should remind that the coefficients are standardized, hence they are comparable to each other in magnitude. In this respect, the negative coefficients are on average larger than the positive ones, which may ease the comprehension of this output. In all events, a deeper look at time evolution of the pandemic can inform better on the role of these factors in explaining the variance of the SWB-I. In fact, if we look at the variable `iDeaths`, we may note that its coefficient is strongly negative in February and also negative in May and July while `iCases` has a negative impact in April and June with a coefficient higher, in absolute terms, than `iDeaths` for the same months (when the effect of `iDeaths` is positive): this indicates that the reactions to the Covid-19 outbreak change over time, following the relative intensity of the phenomena (affections, casualties) and are, to some extent, substitutes. On the other hand, the `iCovid` coefficient is negative in April, May, June and August and `iCoronaVirusNews` is negative from January till April and then again in July and so forth. This means that the news about the pandemic (`iCovid`, `iCoronaVirusNews`, etc) and the effect of the pandemic itself (`iDeaths`, `iCases`, `iRt`) are mostly depressing the value of the well-being up to summer, then some relief seems to appear. Clearly, some of these input are probably substitutes in terms of psychological effects. Another interesting aspect is the `iUnemployment` which seems to be relevant and negative before or at the onset of the pandemic (January and February) and before the start of the second wave (September), showing that, in the acute phase of the pandemic, increasing fears for health crowd out the economic concerns.

[18] For the analysis we used the Spearman correlation coefficient that is a rank-based measure of association. This measure is known to be more robust and it possibly captures also association between non-linear, yet monotonic, transformation of the data.

[19] As the data set contain 27 covariates and only 30 days of observations, we were forced to run a step-wise regression analysis.

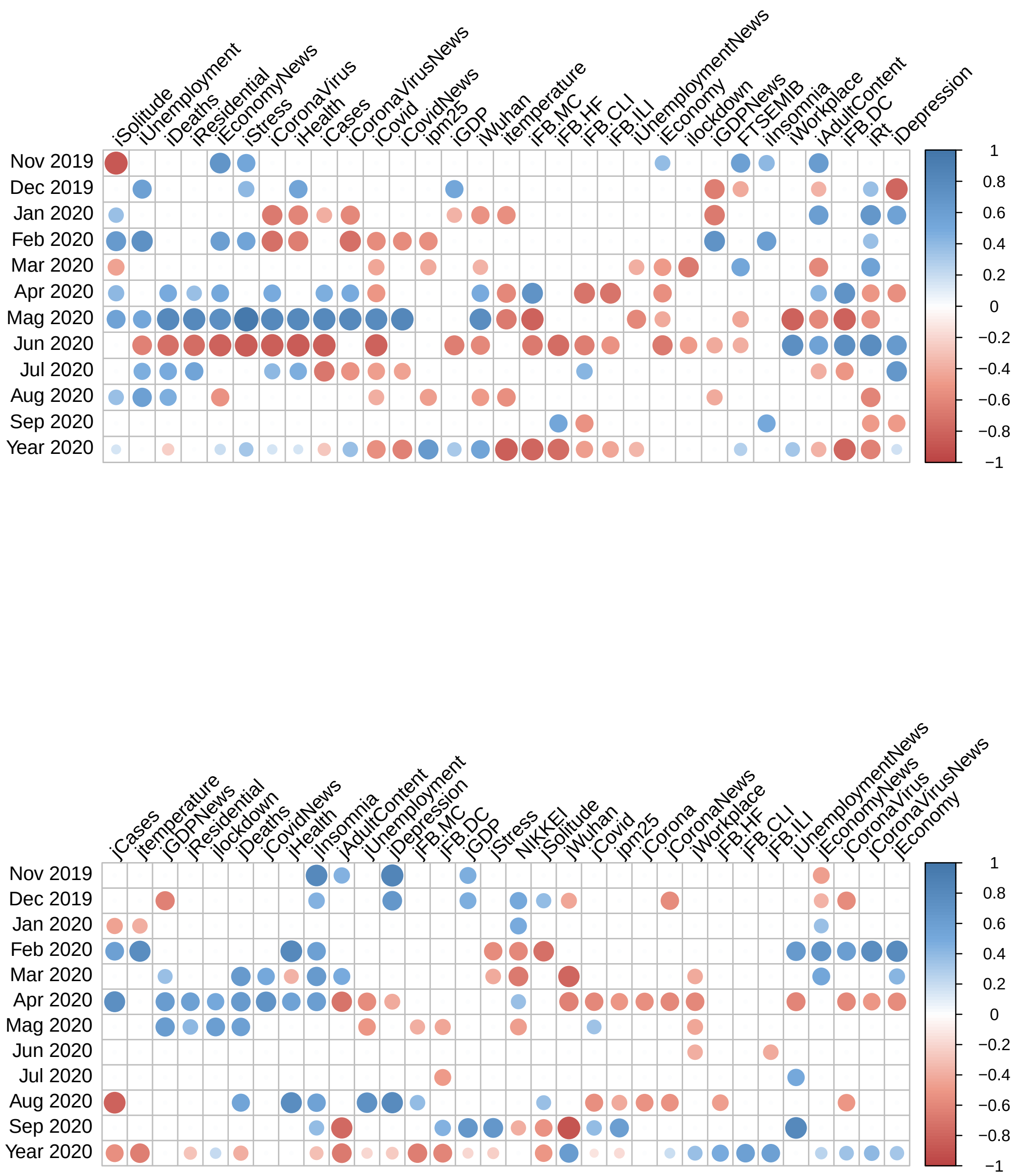

**Figure 2:** Correlation analysis of SWB-I (top panel) and SWB-J (bottom panel) versus all the covariates used in the analysis, by month and for the year 2020 (January to September). Spearman correlation coefficient, only shown statistically significant estimates with $p$-values $<0.05$.

Looking at Japan, the variable `jCases` seems to be more important that `jDeaths` across months, probably because Japan experienced much lower number of fatalities than every other country in the world compared to the number of cases. In particular `jDeaths` shows a positive coefficient, which seems quite natural as the SWB-J decreases through time but the number of deaths is quite low an stable. Overall, for the whole period January to September, the news about coronavirus and Covid-19 seem to have a negative impact as well as the psychological health proxies (`jStress`, `jHealth`, `jSolitude`, `jInsomnia`), the mobility variables (`jWorkplace`, `jResidential`) and some of the economic variables (`jGDPNews`, `jUnemployment`). In more scattered way, this is confirmed through the months, where these variables appears to negatively impact the well-being in different times.

The reader may dig more into the analysis of the monthly regression and provide further interpretation of the coefficients and the impact of the different factors on the SWB indicators. Still, if we accept that the explanatory variables' impact on SWB rapidly changes over time, we should also accept that the calendar month cut may be a bit arbitrary. For this reason, we introduce a dynamic and more thorough approach, based on data science techniques, in the next section.

## 6.3 Dynamic Elastic Net Analysis

Having seen the features and limits of a yearly or monthly based analysis, we now present a dynamic model selection analysis. This means that instead of running a single fit of a statistical model on the whole or monthly data, we consider a sliding window of 30 days. In each 30 days-long time series data, we fit a the same statistical model discussed in the previous section, applying the Elastic Net approach (Zou and Hastie, 2005), which is a regularized estimation method that performs estimation and model selection at the same time[20]. In our setup, the Elastic Net method is a penalized least squares method which adds $L_1$ and $L_2$ penalization terms to the classical objective function of the OLS, i.e., it corresponds to the optimization problem in (5) below:

$$\underset{\beta}{\operatorname{argmin}} \left\{ \frac{1}{2n} \sum_{i=1}^{n} (y_i - x_i'\beta)^2 + \lambda \left( \frac{1-\alpha}{2} \sum_{j=1}^{k} \beta_j^2 + \alpha \sum_{j=1}^{k} |\beta_j| \right) \right\}. \tag{5}$$

where $y_i$ si the dependent variable and $x_i = (x_{i1}, x_{i2}, \dots, x_{ik})'$ is the vector of covariates for unit $i$ in the sample of $n$ observations, $\beta_j$s are the regression coefficients, $\lambda > 0$ is a penalization factor and $\alpha \in [0, 1]$ is a tuning parameter. For $\alpha = 1$ this method corresponds to the classical LASSO algorithm (Tibshirani, 1996), while for $\alpha = 0$ it corresponds to Ridge estimation (Hoerl and Kennard, 1970). Loosely speaking, while the LASSO method tends to estimate as zero as much coefficients as possible and, in case of multicollinearity, selects arbitrarily one single variable in a set of correlated covariates, the Ridge regression is able to accommodate for the multicollinearity by keeping the correlated variables and "averaging" the estimated coefficients. For exactly this reason Elastic Net include both $L_1$ (LASSO) and $L_2$ (Ridge) penalty terms. Usually LASSO is used to succinctly explain the correlation effects and Ridge is more suitable for forecasting. Overall, these methods are biased[21] but - with low variance, as they are also shrinkage methods, and hence most of the time - the Elastic Net performs quite well in terms of mean squared error compared to classical OLS. The value of $\alpha = 0.5$, usually denotes the proper Elastic Net. The penalty term $\lambda$ is a tuning parameter and it is chosen through cross-validation methods. In this study we make use of the package `glmnet` developed by Friedman et al. (2010), which is computationally efficient and state of the art for this technique, to run several times the Elastic Net model. In particular, instead of a simple regression, we estimate a one step ahead forecasting model of this form

$$\underset{\beta}{\operatorname{argmin}} \left\{ \frac{1}{2 \cdot 30} \sum_{d=t-29}^{t} (y_d - x_{d-1}'\beta)^2 + \lambda_t \left( \frac{1-\alpha}{2} \sum_{j=1}^{k} \beta_j^2 + \alpha \sum_{j=1}^{k} |\beta_j| \right) \right\}, \quad t = \text{2020-12-02}, \dots \tag{6}$$

where $\lambda_t$ is calculated for each varying time $t$ through cross-validation minimizing the mean squared error of the forecast. In the set of regressors we also include the lagged value of the index (`swbLag`) and compare the forecasting against an ARIMA(1,0,1) model. We tested the average mean squared error of the forecast against that of the ARIMA(1,0,1) and they are quite close, with the latter being slightly better[22]. Clearly the ARIMA(1,0,1) model does not include any covariate but takes advantage of modeling the serial correlation, whilst Elastic Net in (6) essentially assumes independent observations - obviously a simplification of the reality - but it allows for the inclusion of explanatory variables. Figure 3 shows the relative performance of the Elastic Net compared to the ARIMA(1,0,1) model for both SWB-I and SWB-J. For completeness, we also tested the three values of $\alpha = 0, 0.5, 1$ getting that, as expected, $\alpha = 0.5$ gives the best performance. A final remark: as the real scope of this study is to evaluate the impact of the covariates on the SWB indexes, the actual forecasted values are not important, so we standardize the data as Elastic Net works better with standardized data. Moreover, as the standard error of the estimates is quite difficult to obtain and it is not a typical outcome of the procedure, via the standardized data we also have standardized coefficients which make the evaluation of the impact of each variable easier to interpret (see Figures 5-6). Still, the relative importance of each variable is also obtained applying a Random Forests algorithm and extracting the importance measure from it. Once the importance measure is available, the covariates are ranked accordingly and a relative rank is calculated, where the relative rank is equal to 1 if the rank is highest, and is equal to 0 if the variable has not been selected (see Figures 7-8). Anyway, our main interest is in verifying the patterns of correlation through time. These patterns appears quite clearly indeed, as it will be discussed in the following.

[20] Contrary to the step-wise regression analysis, which is a hierarchical method that does not explore all possible solutions.
[21] Adaptive versions of Elastic Net also exist.
[22] The average MSE is 0.01734827 for Elastic Net and 0.0152734 for ARIMA(1,0,1) for the Italian data and 0.08026962 and 0.06737153 respectively for the Japanese data.

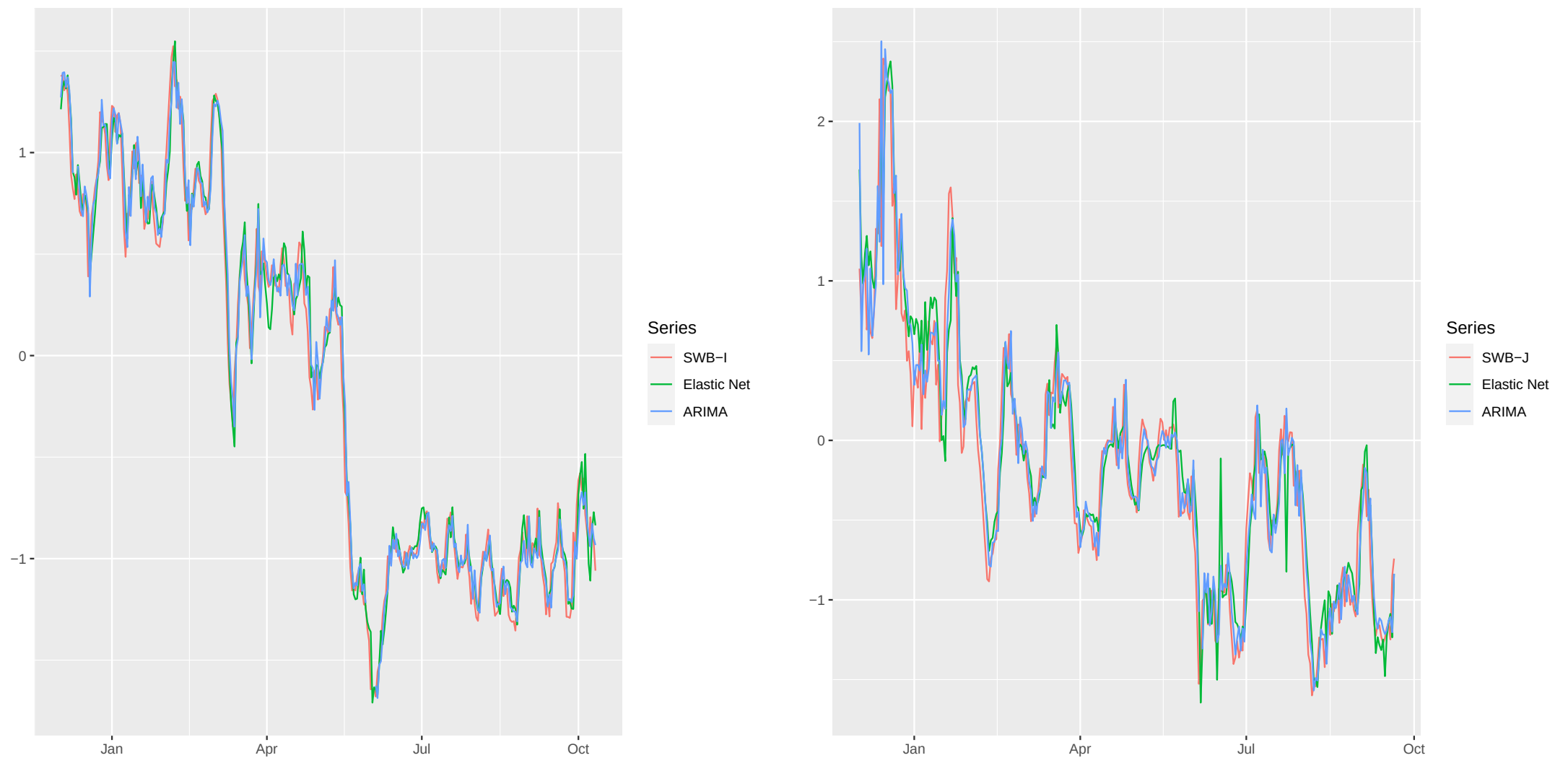

**Figure 3:** Forecast performance of Elastic Net and ARIMA(1,0,1). Standardized data.

## 6.4 Analysis of the Italian Data

By looking at Figure 5, it is clear that the variable `iDeaths` has a negative coefficient from mid February till the end of March (at the peak of the first wave), then again appears isolated on the 25th of May and again at the end of September and early October when the second waves restarted. Interestingly enough, `iCases` started much earlier around the end of January (remind that all flights from China to Italy were stopped on January 21st) and vanishes as `iDeaths` pops up. Then, its coefficient appears again and positive from mid May to mid June, when the good news of the decreasing in number of cases arrived. A similar pattern is exhibited by the `iWuhan` variable. It is worth noting that the `iRt` variable becomes negative by the end of April and quite persistent starting from the summer. This is clearly a media effect, as the notion of $R_t$ itself became known to the public opinion only in July when both opinion makers and virologists started to discuss the indicator in the media, also on prime time tv shows.

The variable `ilockdown` and `iDepression` appear in mid March[23], and the latter is also persistently negative between the end of April and mid May.

The Facebook survey variables `FB.ILI` and `FB.CLI` appear from their beginning (April 27th) and remain almost always negative (apart around the last weeks of July), meaning that the symptoms clearly impacted negatively the well-being.

The variable `iUnemploymentNews` is negative mostly from August, which is when the economic impact on several jobs, including seasonal jobs, had the largest impact and frequently reported in the news. It can be noted, moreover, that the `iUnemployment` effect is much less sharp, suggesting a higher incidence of the labor market perspectives, with respect to the impact of the current employment conditions.

The `itemperature` has a positive impact in the winter (which has been mild in Italy) and negative in the summer, which is somewhat confirmed by common experience, and the air pollution variable `ipm25` has almost always a negative impact when it is significant.

The mobility variables also show interesting patterns: `iResidential` is often positive unless in August, when staying at home was quite challenging, while `iWorkplace` has been negative around the last week of March and the first week of April, then generally positive when the mobility constrained where lifted in most regions.

The `iAdultContent` variable is always negatively correlated with the well-being, confirming that the consumption of these products are closely related to unsatisfactory levels of happiness.

[23]The lockdown in Italy started officially on March 15th

The impact of economy related news and indexes are somehow questionable as they switch sign or are scattered along the time axis. This may indicate that these effects are extremely volatile and mainly determined by the correspondent news, as we may infer from `iGDP` and `iGDPNews`: in fact, while `iGDP` has sometimes a positive sign, the expected negative role seems played by `iGDPNews`. On the other hand, one may think that the impact of the economic factors is a long-run one, and hence is not properly captured by a short-run SWB indicator.

## 6.5 Analysis of the Japanese Data

The first wave of the pandemic started in Japan with the beginning of 2020, while - after a slowdown - a second wave arose in June. Figure 6 shows that the variable `jCases` is quite often present from January till May, then in June and August, and with almost always negative sign, whilst the `jDeaths` variable seems to be selected much less frequently and with alternate sign. This is probably because the number of deaths in Japan is way less than the number of cases. Also, the cases appeared in Japan much earlier than in Italy, and this is reflected by the selection of the variables.

Psychological aspects captured by the Google search `jStress` and `jSolitude` are are mostly present with negative sign while, surprisingly enough, `jInsomnia` and `jDepression` seems to be positively correlated with the SWB-J index, and the latter only up to mid February.

The economic variables - stock market index `NIKKEI`, `jEconomy`, `jFB.HF` and `jGDPNews` - are often selected and with negative impact on subjective well-being: in particular, labor market related covariates (`jUnemployment`, `jUnemploymentNews`) frequently appear, even with alternate sign, to measure concerns for current and future employment perspectives.

As expected, `jAdultContent` is negative when present, and seems to be relevant before the start of the pandemic and in between the two waves mostly. Covid-19 symptoms, as registered through the Facebook survey, seem to be less important apart for the period mid July - mid August when `jFB.CLI` appears with negative sign.

The mobility variables `jWorkplace`, `jResidential` and `jlockdown` appear with negative sign but they are not very persistent in time like it appeared for the Italian case. In fact, the mobility restrictions in Italy were much stronger and this may be the reason of this different impact. On the contrary, the social distancing proxies `jFB.MC` and `jFB.DC` appear and remain persistent during the second wave of the pandemic, and have both a negative impact on the SWB-J index.

The search terms related to Covid or coronavirus are also scattered and with alternate sign, showing a volatile relationship with on well-being, not so different from what one may observe in the Italian case.

The variable `jtemperature` behaves similarly to the Italian case: often positive in winter/spring and negative in June and August, and again positive in the first half of July. The air quality proxy `jpm2.5` is rarely significant compared to the Italian case.

Neither the postponement of the Olympic Games 2020, at the end of March, nor the consequent panic wave - with lines outside the supermarkets - both registered by Twitter and by SWB-J, are associated with the selection of specific covariates. The same happens with the resignation by the prime minister Shinzo Abe (end of August). On the other hand, the well-being reaction at the second wave outbreak in June is accompanied by a selection of all the covariates, some of which (`jCases`, `jCorona`) with a strongly negative impact.

## 6.6 Comparative Analysis of the Elastic Net Results

Figure 4 summarizes the dynamic variable selection analysis for the two countries together. It shows the number of days each predictor has been selected by the elastic net model in equation (6) and the average relative importance of each variable as determined by the Random Forest algorithm after each step of the analysis. This representation is alternative to the one in Figures 7 and 8, as the time progression is not reported, but it still gives a glimpse of the overall impact of each variable along the whole period of analysis. In relative terms, it shows, for example, that the variables `jDeaths` and `jCases` are more frequently selected compared to the corresponding factors `iDeaths` and `iCases`, as well as the fact that the number of deaths is more important - compared to the number of cases - to explain SWB-I, while the number of cases is more important to explain SWB-J.

The `itemperature` and `jtemperature` variables are frequently selected in both countries, but this is not the case for the air quality proxy `ipm2.5` which is way more frequently selected than `jpm2.5` counterpart.

In general, the economic variables are more often selected in the Japanese case: see, for instance, `NIKKEI`, `jEconomyNews`, `jUnemployment` compared to the corresponding `FTSEMIB`, `iEconomyNews` and `iUnemployment` variables. On the contrary, `ilockdown` is more important compared to `jlockdown`, which is in line with the evidence reported in the previous sections and with the more stringent restrictive measures adopted in Italy. Further, the Covid-like symptoms variable `iFB.CLI` is more frequently selected than `jFB.CLI`, and so forth.

All this to emphasize, once again, that the complexity of well-being determinants and their relationships - encompassing cultural dimensions, economic status, environmental conditions as well as many contingent factors - is hardly described by a single model.

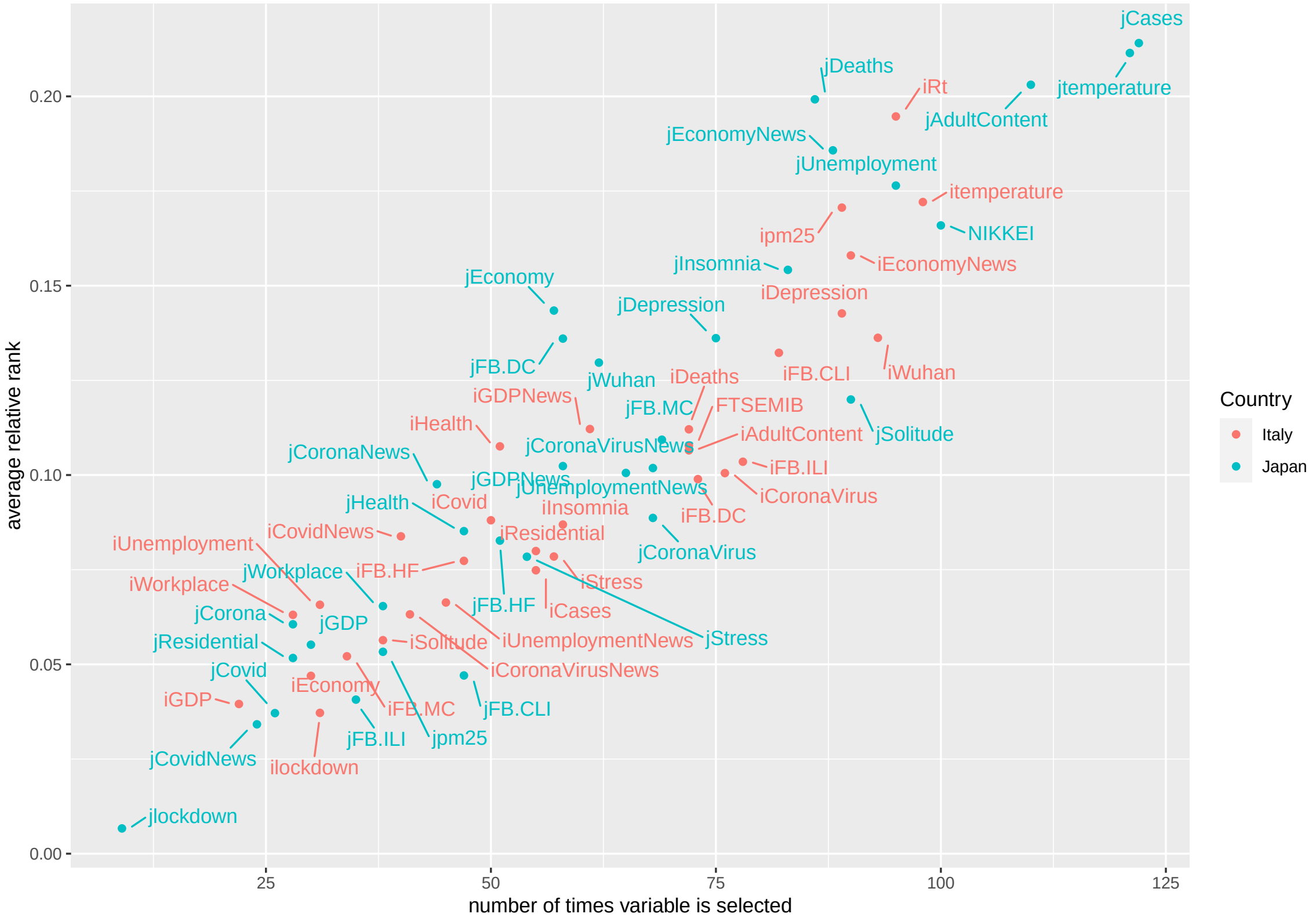

**Figure 4:** Summary of relative importance (average relative rank) and number of times each variable is selected (over 315 dates for Italy and 294 dates for Japan) by the dynamic elastic net.

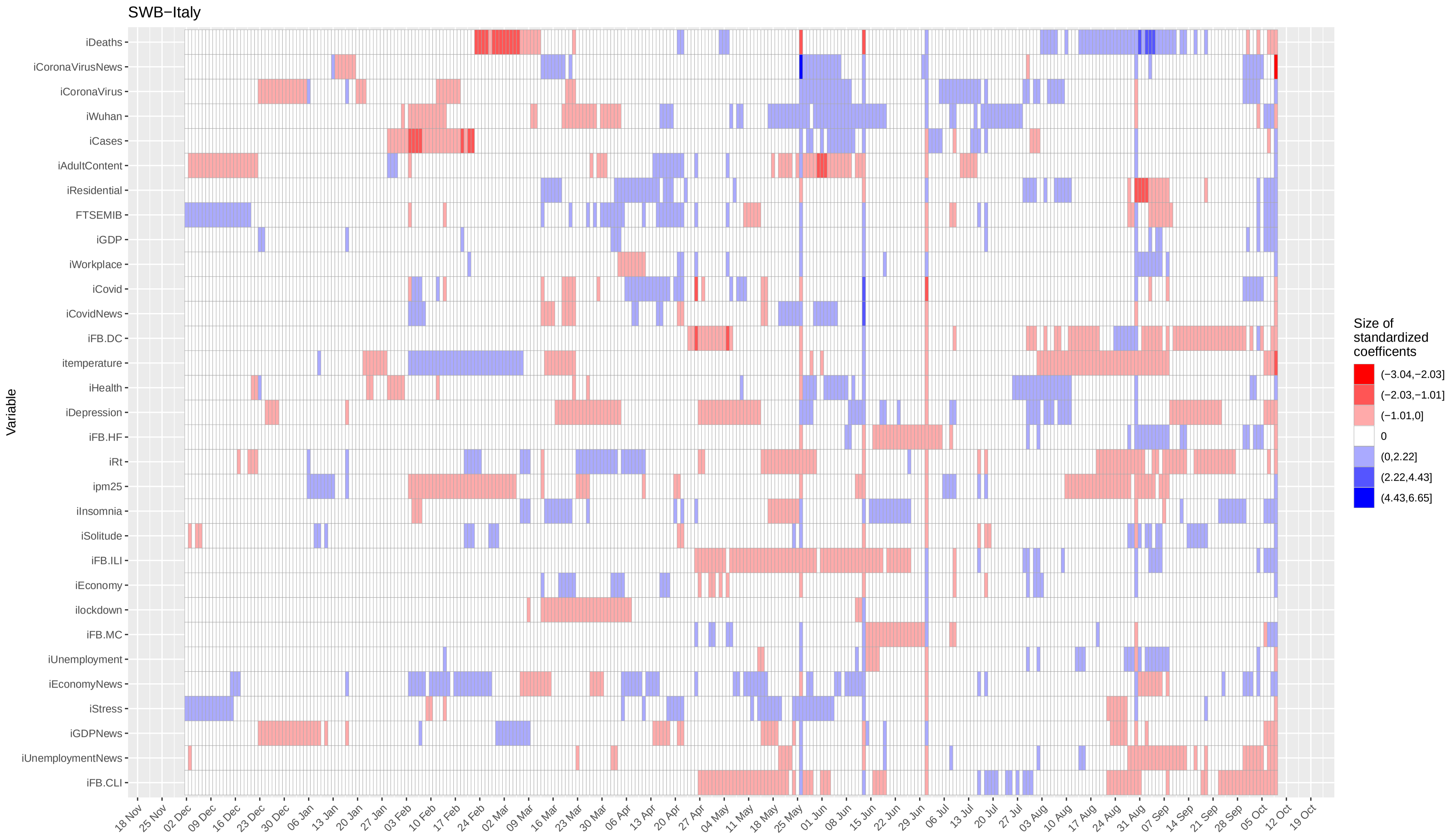

**Figure 5:** Standardized coefficients of the covariates selected by the Elastic Net for SWB-I through time.

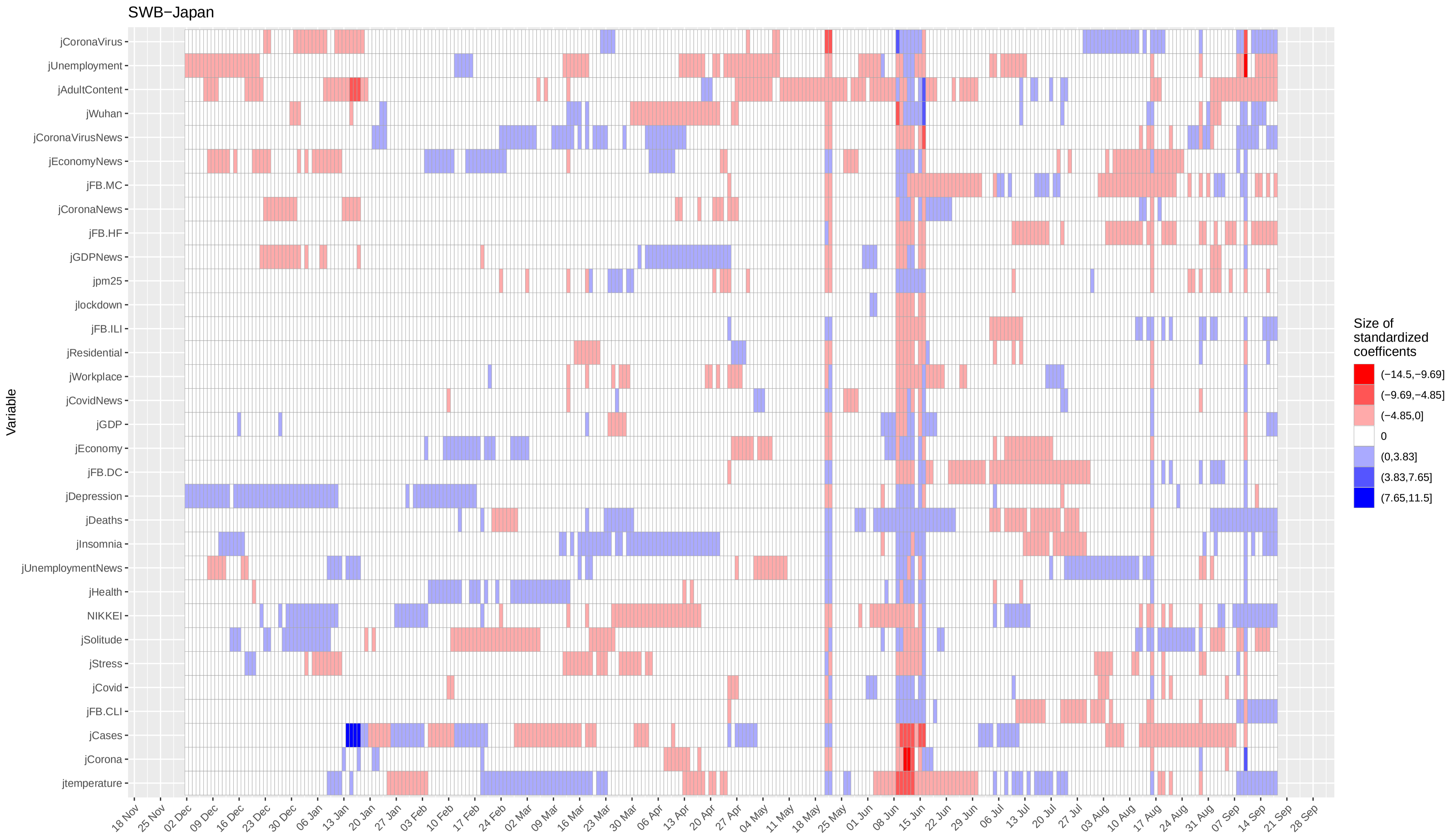

**Figure 6:** Standardized coefficients of the covariates selected by the Elastic Net for SWB-J through time.

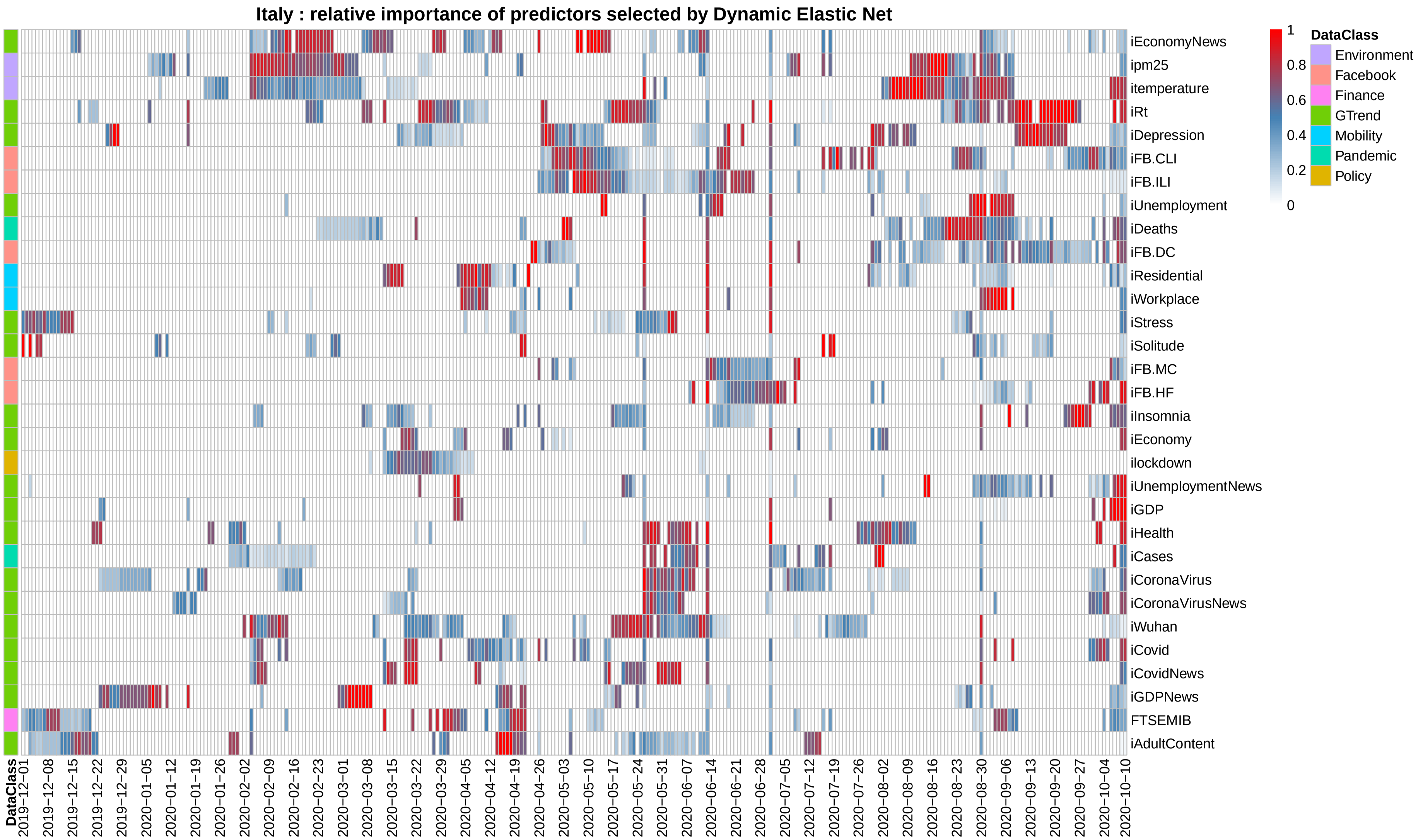

**Figure 7:** Relative importance (1 = maximum, 0= variable not selected) of the covariates selected by Elastic Net to explain the SWB-I through time.

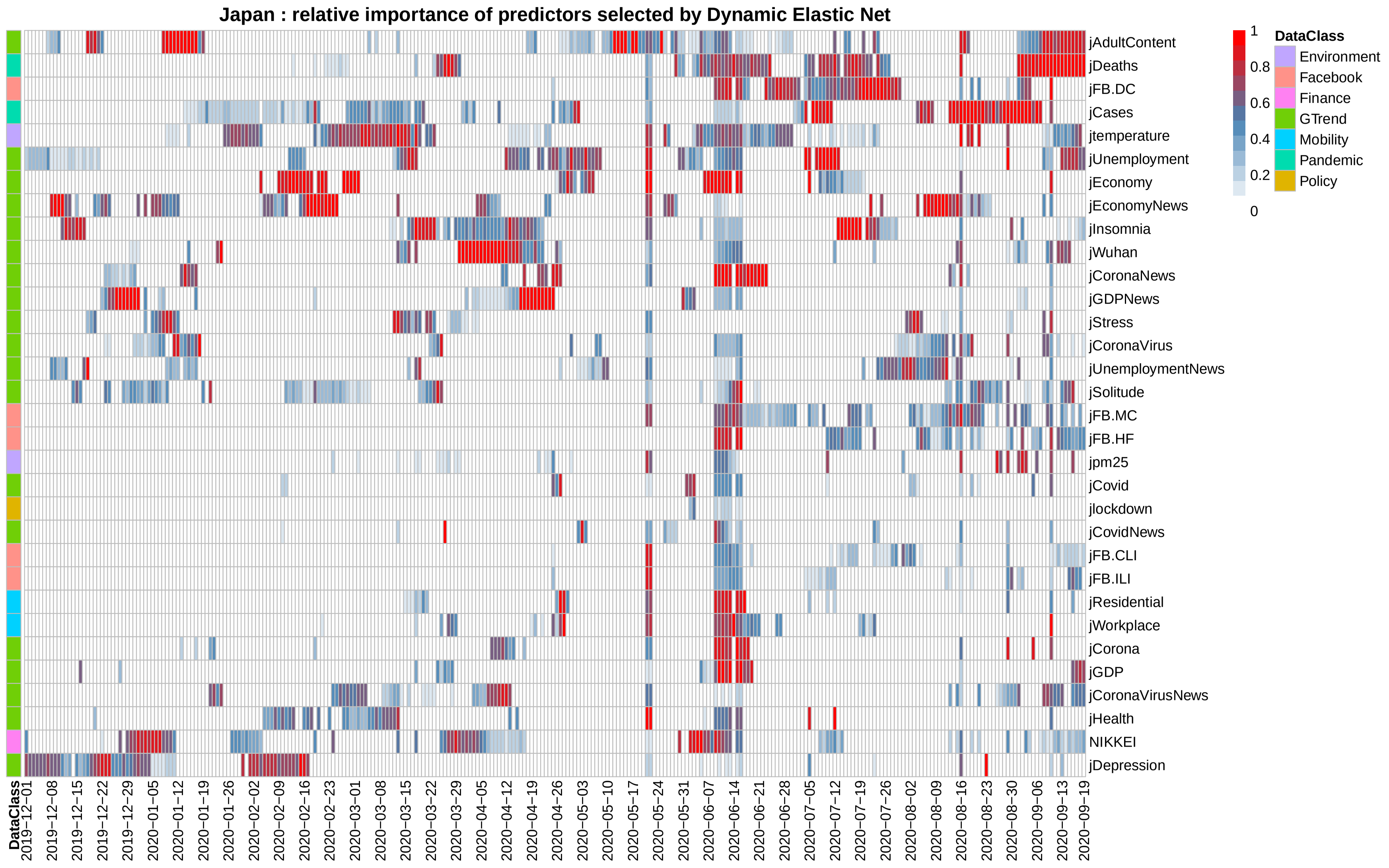

**Figure 8:** Relative importance (1 = maximum, 0= variable not selected) of the covariates selected by elastic Net to explain the SWB-J through time.

# 7 Structural Equation Modeling

As seen in the previous sections, the relationships between the different factors used in this study and the subjective well-being indexes are quite complex and are mediated by an underlying latent variable that we may call subjective well-being. Moreover, some of the covariates are clearly correlated and may be considered as proxies of other latent dimensions. For these reasons, we try to summarize these relationships via a Structural Equation Modeling (SEM) with continuous response variable (Bollen, 1989) approach. SEM is a common method to test complex relationships between dependent variables, independent variables, mediators, and latent dimensions. In statistical terms, SEM consists of regression analysis, factor analysis, and path analysis applied to explore interrelationships between variables. It is a confirmatory technique where an analyst tests a model to check consistency between the relationships put in place.

We assume that well-being is a latent variable influenced by external factors. Further, the following latent dimensions, whose interpretation is postponed after the analysis of the results (see also Table 7), are assumed to exist:

- `VirusSearch`: captures the impact on `Well-being` of Covid-related information web search. We assume the latent variable to be determined by the following observable covariates: `CoronaVirus`, `Covid` and, only for Italy by `Rt` and, only for Japan by `Corona`;
- `PsySearch`: captures the impact on `Well-being` of web search for `Stress`, `Insomnia`, `Solitude` and `Depression`;
- `HealthStatus`: captures the declared health status, whose observables are `FB.CLI` and `FB.ILI`.
- `Mobility`: captures the mobility restrictions factor, measured through the variables `Residential`, `Workplace` and `lockdown`;
- `Finance`: accounts for the effect of the financial and economic variables: `FTSEMIB/NIKKEI`, `FB.HF` and `Unemployment`.
- `SocDist`: represents the social distancing factor, measured by `FB.MC` and `FB.DC`.

and the main causal relationships are as follows:

$$
\begin{aligned}
\text{Well-being} &\leftarrowtail \text{VirusSearch} + \text{HealthStatus} + \text{Mobility} + \text{Finance} + \text{SocDist}\\
\text{PsySearch} &\leftarrowtail \text{Well-being}\\
\text{iAdultContent} &\leftarrowtail \text{Well-being}\\
\text{SWB-J/SWB-J} &\leftarrowtail \text{Well-being}
\end{aligned}
$$

Further, the residual correlations among some of the observed variables are inserted in the model to take into account cross-correlation, for example between `Mobility` and `Cases`. The results of the fitted model are displayed in Tables 10-11, while Figures 10-11 give a graphical representation of the same fitting. In light of the results of the SEM models, Table 7 is an attempt to explain the meaning of the latent dimensions introduced in the analysis. The models have been fitted using the `lavaan` package (Rosseel, 2012) and plots have been generated through the `semPlot` package (Epskamp, 2019).

## 7.1 Comments to the Structural Equation Modeling Analysis

Once established that the sign, the significance and the relative importance of observable variables in determining subjective well-being tend to rapidly change, we may try to derive from the SEM analysis a synthetic picture of the connections between the observables and well-being, with the help of latent dimensions.

Figures 10-11 show that the latent variables affect well-being with the expected sign, given the interpretation of these variables provided in Table 7: social distance measure and concerns, a worse confidence in economic and financial conditions, covid-like and flu symptoms, mobility restrictions and web search for Covid-related terms negatively impact well-being. We should note, however, that the `Finance` latent variable does not exhibit a significant effect on Japanese well-being, probably due to less severe concerns about the economic consequences of the pandemic, compared to the Italian case: we must remember that the estimated GDP reduction in 2020 in Japan is around a half with respect to what is expected in Italy[24]. Figure 9 shows the real GDP growth in 2020 and the number of COVID-19 deaths per 1 million inhabitants, and it is quite informative of the relative impact of COVID-19 on the different economies around the world.

A partial exception is represented by the latent variable `PsySearch`, that it is supposed to summarize the psychological status measured by the web search of terms related to anxiety, loneliness and depression: while this variable is negative impacted by well-being in Japan, this does not happen for the Italian case, where the relationship is surprisingly positive and significant: this is one of the cases where cultural habits and social norms may play a role in producing different results in different countries.

Finally, we have to point out the positive correlation between the two SWB indexes and the corresponding well-being latent variable. On one hand, this pleads for the ability of the indicators to read into public opinion;

[24]The estimated real GDP growth in 2020 is -10.6% for Italy and -5.3% for Japan according to the International Monetary Fund. The full report can be found at: https://www.imf.org/external/datamapper/NGDP_RPCH@WEO/OEMDC/ADVEC/WEOWORLD?year=2020

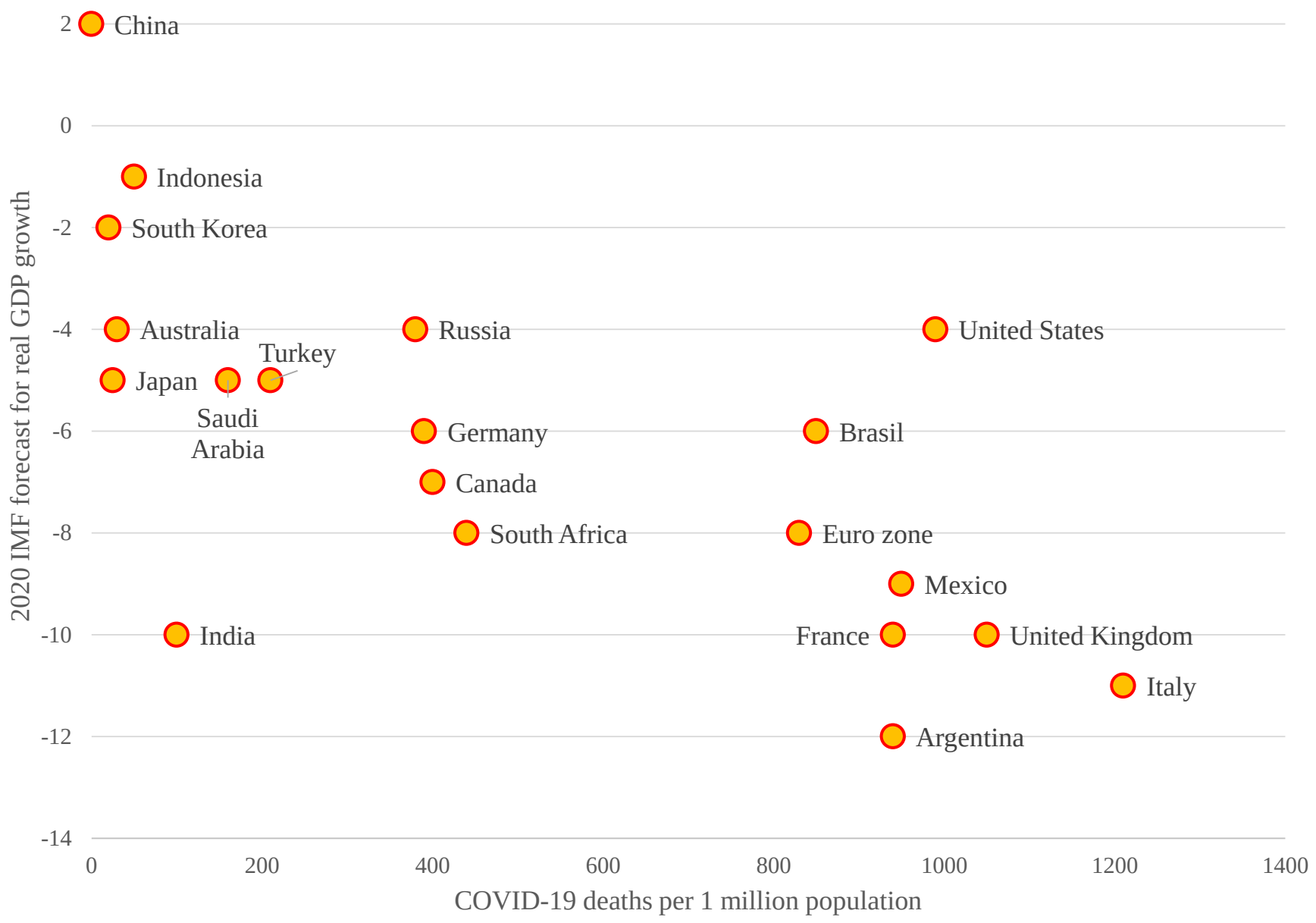

**Figure 9:** Number of COVID-19 deaths (WHO) and real GDP forecast growth in 2020 (IMF).

on the other hand, we cannot help noting that the relationship is much stronger for SWB-J, suggesting that the Twitter-based tool better capture the individual and collective sentiment in the Japanese context.

| Latent | Country | Interpretation | Relationship |
|---|---|---|---|
| `VirusSearch` | IT | *pandemic getting better* | the higher its value, the lower the web search for reproduction number $R_t$ and Covid in general |
| | JP | *fear about the pandemic* | the higher its value, the higher the web search for Corona virus and Covid in general |
| `HealthStatus` | IT, JP | *fear about symptoms* | the higher its value, the higher concerns about health |
| `Mobility` | IT, JP | *mobility restrictions* | the stronger its implementation, the harder to stand restrictions |
| `SocDist` | IT, JP | *social distancing practice* | the wider its use, the harder to stand distancing |
| `Finance` | IT | *fear for own economic conditions* | the higher its value, the higher the pressure on own life |
| | JP | *confidence in country economy* | the higher its value, the better |
| `PsySearch` | IT | *search for psychological symptoms* | mixed evidence, hard to conclude |
| | JP | *psychological status* | the higher its value, the worse the well-being |

**Table 7:** Ex-post interpretation of the latent variables of the SEM model. The `VirusSearch` and `Finance` convey similar information but with switched signs in both countries. The rest of the latent variables have the same and intuitive interpretation apart from `PsySearch` which has apparently no clear interpretation for the Italian case.

# 8 Results and Discussion

The main evidence of this study is that a single approach cannot give a full account about the complex relationship among subjective well-being and the wide range of potential explanatory variables, when an unprecedented shock hits the socio-economic context. Nevertheless, the multi-method analysis conducted in the previous sections let a few important evidences emerge:

- the Twitter indexes SWB-I and SWB-J show that Italy and Japan have a substantial difference in their

own self-perceived well-being: the Italian indicator is permanently higher than the Japanese one. On the other hand, both SWB-I and SWB-J report a substantial decrease in 2020 compared to the previous years (-11.7% for Italy and -8.3% for Japan);

- applying a stochastic differential equations analysis, it can be seen that this drop has stabilized since April/May for both countries, meaning that the pandemic shock is provoking a prolonged status of emergency and stress, with permanent negative impact on the self-perceived well-being trend;
- the stochastic analysis also allows for disentangling long term (or structural) and short term (or emotional) components of the subjective well-being, showing that, if the well-being decrease in 2020 is higher in Italy, the convergence to the new long term well-being level is faster in Japan. Implicitly, this analysis clarifies why yearly survey may lead to biased evidence of the real well-being level of a country;
- both the SWB-I and SWB-J Twitter indexes, despite the cultural differences of the two countries, seems to be able to capture some aspects of subjective well-being and the relationships with potential explanatory variables. This ability is not affected by language differences and specificities but, not surprisingly, requires a dynamic analysis to emerge: one of the main advantages of a Twitter-based index, in fact, is in its high frequency and in the opportunity to capture even rapid changes in the relative importance of determinants;
- in fact, the Dynamic Elastic Net analysis has captured the complexity of the well-being determination: which factors matter more than others and when. Although a general model of well-being is desirable, the analysis has shown that according to different context conditions (pre-pandemic, pandemic, in-between waves of pandemic, etc) some determinants weight more than others and they interact differently with cultural aspects in the two countries;
- even if we think to have documented that “average” effects may be misleading guides in interpreting dynamic phenomena, the structural equation modeling analysis has shown some common latent dimensions that can help in describing a causal model for the latent variable of subjective well-being of which the Twitter indexes SWB-I and SWB-J seem to be measurable proxies. In particular, social distances concerns, mobility restrictions, economic conditions, along with more obvious health conditions and Covid-related concerns, have been identified as possible explanatory variables. At the same time, variables linked to psychological dimensions show less expected effects, indicating the need for a more thorough understanding of cultural specificities of the examined countries.

| | *Dependent variable:* | | | | | | | | | |
|---|---|---|---|---|---|---|---|---|---|---|
| | SWB-I | | | | | | | | | |
| | Jan-Sep | Jan | Feb | Mar | Apr | May | Jun | Jul | Aug | Sep |
| iDeaths | | | −98.06$^{**}$ | | 0.62$^{*}$ | −1.67$^{***}$ | | −6.57$^{***}$ | | |
| | | | (39.87) | | (0.32) | (0.25) | | (2.13) | | |
| iCases | | | 15.76 | | −1.35$^{***}$ | 1.34$^{***}$ | −2.44$^{***}$ | | 1.91$^{***}$ | −2.39$^{*}$ |
| | | | (10.39) | | (0.35) | (0.39) | (0.61) | | (0.65) | (1.30) |
| FTSEMIB | | −0.39$^{*}$ | 0.68$^{***}$ | 0.31$^{***}$ | 0.23 | 0.19 | −0.20$^{***}$ | | 0.20 | |
| | | (0.23) | (0.23) | (0.10) | (0.20) | (0.16) | (0.06) | | (0.17) | |
| iCoronaVirus | | | 5.11$^{*}$ | −3.81$^{***}$ | 162.18$^{**}$ | 13.58$^{***}$ | 23.87$^{***}$ | 8.88$^{***}$ | −49.04$^{**}$ | −24.12$^{***}$ |
| | | | (2.90) | (0.62) | (55.84) | (2.15) | (2.99) | (2.01) | (18.71) | (7.31) |
| iCoronaVirusNews | 0.19$^{***}$ | −0.84$^{**}$ | −13.02$^{*}$ | −2.03 | −146.85$^{**}$ | 8.05$^{**}$ | −12.51$^{***}$ | −3.15$^{***}$ | 50.25$^{***}$ | 37.32$^{***}$ |
| | (0.05) | (0.32) | (6.69) | (1.24) | (51.55) | (2.91) | (2.25) | (0.40) | (17.05) | (7.07) |
| iCovid | 0.09$^{***}$ | | 47.16$^{*}$ | 41.20$^{***}$ | −93.24$^{**}$ | −3.18$^{***}$ | | | −4.91$^{**}$ | 2.87$^{***}$ |
| | (0.03) | | (25.16) | (9.94) | (35.02) | (0.89) | | | (2.26) | (0.92) |
| iRt | −0.05$^{***}$ | −0.96$^{**}$ | −12.90 | | 8.29$^{**}$ | 0.13$^{**}$ | 0.73 | | 1.00$^{*}$ | 3.23$^{***}$ |
| | (0.01) | (0.41) | (9.99) | | (3.13) | (0.06) | (0.46) | | (0.49) | (0.82) |
| iCovidNews | | | | −68.92$^{***}$ | −59.26$^{***}$ | | −6.22$^{***}$ | | 5.55$^{**}$ | −5.30$^{***}$ |
| | | | | (17.28) | (19.71) | | (1.11) | | (2.16) | (1.00) |
| iUnemployment | 0.05$^{***}$ | −0.41$^{**}$ | −19.96$^{*}$ | 1.22$^{***}$ | 0.99$^{***}$ | 1.24$^{*}$ | −1.47$^{**}$ | | | −8.90$^{***}$ |
| | (0.02) | (0.18) | (11.06) | (0.28) | (0.33) | (0.70) | (0.58) | | | (1.97) |
| iGDP | −0.11$^{***}$ | | | | | | | | | |
| | (0.03) | | | | | | | | | |
| iGDPNews | 0.03$^{***}$ | | | | | | | | | |
| | (0.01) | | | | | | | | | |
| iHealth | −0.19$^{***}$ | | | | | | | | | |
| | (0.05) | | | | | | | | | |
| iInsomnia | −0.02 | | | | | | | | | |
| | (0.02) | | | | | | | | | |
| iResidential | 0.18$^{***}$ | | 3.73 | | | −0.90$^{***}$ | −1.70$^{***}$ | | −2.11$^{***}$ | 4.17$^{**}$ |
| | (0.05) | | (2.80) | | | (0.31) | (0.55) | | (0.69) | (1.50) |
| iWorkplace | 0.16$^{***}$ | | 5.27 | −0.57 | | | | | 0.39$^{*}$ | −3.54$^{**}$ |
| | (0.06) | | (4.29) | (0.36) | | | | | (0.22) | (1.41) |
| ipm25 | | | | −0.57 | −0.80$^{*}$ | −0.56$^{**}$ | −0.26 | | 1.42$^{**}$ | |
| | | | | (0.36) | (0.40) | (0.20) | (0.25) | | (0.58) | |
| itemperature | 0.11$^{***}$ | | 0.70$^{**}$ | 0.98 | 2.41$^{**}$ | | −1.04$^{***}$ | −1.46$^{***}$ | −3.74$^{***}$ | 2.21$^{**}$ |
| | (0.04) | | (0.28) | (0.62) | (0.99) | | (0.34) | (0.27) | (1.14) | (1.00) |
| iWuhan | −0.05$^{***}$ | | | | | | | | | |
| | (0.01) | | | | | | | | | |
| ilockdown | −0.07$^{***}$ | | | −0.11$^{*}$ | | | 0.08$^{***}$ | | | |
| | (0.02) | | | (0.05) | | | (0.02) | | | |
| iFB.CLI | −0.06$^{***}$ | | | | −0.84$^{**}$ | 0.04 | | | | 0.42$^{*}$ |
| | (0.02) | | | | (0.34) | (0.03) | | | | (0.20) |
| iFB.ILI | 0.04$^{*}$ | | | | 1.03$^{**}$ | | | 0.13$^{***}$ | | −0.45$^{**}$ |
| | (0.02) | | | | (0.35) | | | (0.03) | | (0.18) |
| iFB.MC | | | | | 9.49$^{***}$ | 2.43$^{***}$ | 0.25 | 0.76$^{***}$ | | |
| | | | | | (2.96) | (0.34) | (0.23) | (0.23) | | |
| iFB.DC | −0.27$^{***}$ | | | | −64.48$^{**}$ | | 1.68$^{**}$ | | 1.78 | 5.27$^{**}$ |
| | (0.05) | | | | (23.61) | | (0.75) | | (1.25) | (1.98) |
| iUnemploymentNews | | 0.21$^{**}$ | | | | | | | 1.23$^{***}$ | |
| | | (0.09) | | | | | | | (0.39) | |
| iFB.HF | | | | | | −0.03 | −0.31$^{**}$ | | 0.63$^{**}$ | |
| | | | | | | (0.02) | (0.12) | | (0.26) | |
| swbiLag | 0.82$^{***}$ | 0.71$^{***}$ | | 0.30$^{*}$ | 0.38$^{*}$ | −0.40$^{***}$ | −0.27$^{*}$ | | −0.25 | −0.30 |
| | (0.03) | (0.14) | | (0.17) | (0.20) | (0.13) | (0.14) | | (0.15) | (0.19) |
| Observations | 285 | 31 | 29 | 31 | 30 | 31 | 30 | 31 | 31 | 30 |
| $R^2$ | 0.99 | 0.99 | 1.00 | 0.99 | 0.99 | 1.00 | 1.00 | 1.00 | 1.00 | 1.00 |
| Adjusted $R^2$ | 0.99 | 0.98 | 0.99 | 0.98 | 0.97 | 1.00 | 1.00 | 1.00 | 1.00 | 0.99 |
| *Note:* | $^{*}$p<0.1; $^{**}$p<0.05; $^{***}$p<0.01 | | | | | | | | | |

**Table 8:** Monthly step-wise regression analysis for SWB-I from 1 January 2020 to 30 September 2020.

| | Dependent variable: | | | | | | | | | |
|---|---|---|---|---|---|---|---|---|---|---|
| | SWB-J | | | | | | | | | |
| | Jan-Sep | Jan | Feb | Mar | Apr | May | Jun | Jul | Aug | Sep |
| jDeaths | 0.07** | 3.16** | 6.22* | 1.01 | 0.12 | | | −3.03 | 1.02*** | |
| | (0.03) | (1.15) | (3.43) | (0.76) | (0.07) | | | (1.86) | (0.20) | |
| jCases | −0.15*** | | −23.28** | | −1.21 | | 12.77*** | 2.17* | −0.48** | 2.25 |
| | (0.04) | | (9.76) | | (0.83) | | (2.07) | (1.16) | (0.22) | (1.83) |
| NIKKEI | | | | | −0.40** | −0.33* | 0.54** | | | −1.75 |
| | | | | | (0.15) | (0.16) | (0.21) | | | (1.05) |
| jCoronaVirus | −0.16*** | | | 1.54** | 2.31*** | −4.64*** | −7.26*** | −210.31*** | | 29.09 |
| | (0.05) | | | (0.59) | (0.68) | (1.38) | (1.64) | (69.95) | | (21.39) |
| jCoronaVirusNews | −0.07*** | 0.91** | 2.75** | 2.71*** | −10.60** | | | 437.37*** | −21.26*** | 3.74 |
| | (0.03) | (0.33) | (1.09) | (0.76) | (3.93) | | | (134.31) | (5.85) | (1.76) |
| jCorona | 0.17*** | | 12.04 | 6.11*** | 3.41*** | −11.03*** | −121.00*** | −123.40*** | −24.77*** | −14.04 |
| | (0.05) | | (7.69) | (2.12) | (1.02) | (2.06) | (24.43) | (35.52) | (6.51) | (11.97) |
| jCovidNews | 0.04 | | 1.78 | 1.57*** | | | | 62.57*** | 4.58*** | |
| | (0.03) | | (1.50) | (0.40) | | | | (19.09) | (1.18) | |
| jUnemployment | −0.12** | | | | | | | | −3.76*** | |
| | (0.06) | | | | | | | | (1.21) | |
| jUnemploymentNews | 0.06*** | | | | | | | | | |
| | (0.02) | | | | | | | | | |
| jEconomy | 0.09** | | | | | | | | | |
| | (0.05) | | | | | | | | | |
| jGDPNews | −0.03* | | | | | | | | | |
| | (0.02) | | | | | | | | | |
| jStress | −0.07*** | | | | | | | | | |
| | (0.02) | | | | | | | | | |
| jDepression | 0.13*** | | | | | | | | | |
| | (0.04) | | | | | | | | | |
| jHealth | −0.07* | | | | | | | | | |
| | (0.04) | | | | | | | | | |
| jSolitude | −0.08*** | | | | | | | | | |
| | (0.02) | | | | | | | | | |
| jInsomnia | −0.09** | | | | | | | | | |
| | (0.04) | | | | | | | | | |
| jResidential | −0.20** | | 0.92 | | | −0.81*** | −4.72*** | | | −4.23 |
| | (0.10) | | (0.61) | | | (0.19) | (1.31) | | | (2.76) |
| jWorkplace | −0.11 | | | | −0.87** | | −4.03** | 0.73** | | −4.80 |
| | (0.07) | | | | (0.32) | | (1.48) | (0.29) | | (2.36) |
| jtemperature | 0.18* | | 1.26*** | | −1.12** | −2.87*** | −4.54*** | 6.85** | −3.94*** | 4.56 |
| | (0.09) | | (0.42) | | (0.43) | (0.81) | (1.12) | (3.16) | (1.07) | (2.63) |
| jlockdown | | | | | −0.09* | −0.06 | | | | |
| | | | | | (0.05) | (0.04) | | | | |
| jFB.CLI | 0.20** | | | | −0.53** | | 1.12** | −1.06* | −0.88* | −1.51 |
| | (0.10) | | | | (0.20) | | (0.47) | (0.50) | (0.45) | (0.89) |
| jFB.ILI | −0.20*** | | | | 0.67** | | −0.85** | 0.67 | 0.58 | 0.87 |
| | (0.08) | | | | (0.23) | | (0.31) | (0.39) | (0.38) | (0.72) |
| jFB.MC | | | | | −1.00* | 0.95*** | | −1.73 | | 4.30 |
| | | | | | (0.49) | (0.30) | | (1.33) | | (2.08) |
| jFB.DC | −0.15** | | | | 0.81*** | −1.28*** | | −1.68** | | 1.58 |
| | (0.06) | | | | (0.22) | (0.24) | | (0.65) | | (1.70) |
| jFB.HF | 0.10*** | | | | | | −2.48*** | 0.86 | | −1.13 |
| | (0.03) | | | | | | (0.66) | (0.69) | | (1.20) |
| jpm25 | | | | 0.31*** | | 0.49*** | | −1.25*** | −0.16** | 0.93 |
| | | | | (0.10) | | (0.11) | | (0.38) | (0.07) | (0.47) |
| jCoronaNews | | −2.94*** | −14.29* | −8.53*** | −10.24*** | 12.42*** | 90.80*** | 400.19*** | 31.03*** | 10.26 |
| | | (1.00) | (7.91) | (3.01) | (3.15) | (2.53) | (18.39) | (119.65) | (7.58) | (14.80) |
| jCovid | | | 11.59** | −9.88*** | 31.64** | −5.43*** | −41.87*** | −11.54*** | −0.95*** | |
| | | | (4.94) | (2.74) | (14.47) | (1.64) | (8.55) | (3.45) | (0.23) | |
| swbjLag | 0.77*** | 0.72*** | 0.64*** | | 0.43** | 0.76*** | | | −0.19 | −0.58 |
| | (0.03) | (0.11) | (0.21) | | (0.20) | (0.21) | | | (0.17) | (0.50) |
| Observations | 264 | 31 | 29 | 31 | 30 | 31 | 30 | 31 | 31 | 20 |
| $R^2$ | 0.94 | 0.87 | 0.96 | 0.92 | 0.98 | 0.95 | 0.99 | 0.93 | 1.00 | 1.00 |
| Adjusted $R^2$ | 0.93 | 0.86 | 0.94 | 0.89 | 0.96 | 0.92 | 0.99 | 0.86 | 0.99 | 0.99 |

*Note:* $^{*}p<0.1$; $^{**}p<0.05$; $^{***}p<0.01$

| Relationship | | | Coefficient | Std.Err. |
|---|---|---|---|---|
| VirusSearch | $\mapsto$ | iCoronaVirus | 0.226*** | 0.043 |
| VirusSearch | $\mapsto$ | iCovid | −0.638*** | 0.063 |
| VirusSearch | $\mapsto$ | iRt | −0.372*** | 0.057 |
| PsySearch | $\mapsto$ | iStress | 0.602*** | 0.053 |
| PsySearch | $\mapsto$ | iInsomnia | 0.410*** | 0.055 |
| PsySearch | $\mapsto$ | iSolitude | 1.003*** | 0.047 |
| PsySearch | $\mapsto$ | iDepression | −0.186*** | 0.056 |
| HealthStatus | $\mapsto$ | iFB.CLI | 0.929*** | 0.047 |
| HealthStatus | $\mapsto$ | iFB.ILI | 1.010*** | 0.044 |
| Mobility | $\mapsto$ | iResidential | 0.791*** | 0.033 |
| Mobility | $\mapsto$ | iWorkplace | −0.726*** | 0.039 |
| Mobility | $\mapsto$ | ilockdown | 0.762*** | 0.036 |
| Finance | $\mapsto$ | FTSEMIB | −0.778*** | 0.052 |
| Finance | $\mapsto$ | iFB.HF | 0.521*** | 0.053 |
| Finance | $\mapsto$ | iUnemployment | 0.421*** | 0.051 |
| SocDist | $\mapsto$ | iFB.MC | 0.882*** | 0.047 |
| SocDist | $\mapsto$ | iFB.DC | 0.976*** | 0.044 |
| WellBeing | $\mapsto$ | SWB-I | 0.248*** | 0.019 |
| WellBeing | $\leftarrow\!\shortmid$ | VirusSearch | 0.307*** | 0.075 |
| WellBeing | $\leftarrow\!\shortmid$ | HealthStatus | −0.167*** | 0.085 |
| WellBeing | $\leftarrow\!\shortmid$ | Mobility | −0.290*** | 0.124 |
| WellBeing | $\leftarrow\!\shortmid$ | Finance | −0.524*** | 0.101 |
| WellBeing | $\leftarrow\!\shortmid$ | SocDist | −3.345*** | 0.336 |
| PsySearch | $\leftarrow\!\shortmid$ | WellBeing | 0.065*** | 0.016 |
| iAdultContent | $\leftarrow\!\shortmid$ | WellBeing | −0.136*** | 0.016 |
| PsySearch | cov | Mobility | 0.003 | 0.008 |
| Mobility | cov | SocDist | −0.412*** | 0.049 |
| Mobility | cov | iCases | 0.356*** | 0.034 |
| VirusSearch | cov | Mobility | −0.293*** | 0.081 |
| VirusSearch | cov | Finance | −1.227*** | 0.091 |
| VirusSearch | cov | SocDist | −1.089*** | 0.072 |
| HealthStatus | cov | Mobility | −0.002 | 0.057 |
| HealthStatus | cov | Finance | 0.230*** | 0.069 |
| HealthStatus | cov | SocDist | 0.314*** | 0.055 |
| Mobility | cov | Finance | 0.850*** | 0.033 |
| Finance | cov | SocDist | 0.305*** | 0.069 |
| PsySearch | cov | iAdultContent | 0.608*** | 0.046 |

*Note:* *p<0.1; **p<0.05; ***p<0.01

**Table 10:** The fitting results of the SEM model for the Italian data.

| Relationship | | | Coefficient | Std.Err. |
|---|---|---|---|---|
| VirusSearch | ↦ | jCoronaVirus | 0.867*** | 0.050 |
| VirusSearch | ↦ | jCovid | 0.204*** | 0.061 |
| VirusSearch | ↦ | jCorona | 0.714*** | 0.054 |
| PsySearch | ↦ | jStress | 0.401*** | 0.055 |
| PsySearch | ↦ | jInsomnia | 0.749*** | 0.049 |
| PsySearch | ↦ | jSolitude | 0.763*** | 0.049 |
| PsySearch | ↦ | jDepression | 0.545*** | 0.052 |
| HealthStatus | ↦ | jFB.CLI | 1.016*** | 0.042 |
| HealthStatus | ↦ | jFB.ILI | 0.947*** | 0.045 |
| Mobility | ↦ | jResidential | 1.071*** | 0.039 |
| Mobility | ↦ | jWorkplace | −0.902*** | 0.046 |
| Mobility | ↦ | jlockdown | 0.708*** | 0.050 |
| Finance | ↦ | NIKKEI | 0.836*** | 0.050 |
| Finance | ↦ | jFB.HF | −0.441*** | 0.054 |
| Finance | ↦ | jUnemployment | −0.694*** | 0.053 |
| SocDist | ↦ | jFB.MC | 0.959*** | 0.044 |
| SocDist | ↦ | jFB.DC | 0.903*** | 0.047 |
| WellBeing | ↦ | SWB-J | 0.674*** | 0.036 |
| WellBeing | ↤ | VirusSearch | −0.368*** | 0.134 |
| WellBeing | ↤ | HealthStatus | −0.617*** | 0.240 |
| WellBeing | ↤ | Mobility | −0.077 | 0.121 |
| WellBeing | ↤ | Finance | 0.052 | 0.165 |
| WellBeing | ↤ | SocDist | −1.795*** | 0.298 |
| PsySearch | ↤ | WellBeing | −0.342*** | 0.052 |
| jAdultContent | ↤ | WellBeing | −0.456*** | 0.039 |
| PsySearch | cov | Mobility | 0.002 | 0.023 |
| Mobility | cov | SocDist | 0.087 | 0.054 |
| Mobility | cov | jCases | 0.235*** | 0.030 |
| VirusSearch | cov | HealthStatus | 0.536*** | 0.049 |
| VirusSearch | cov | Mobility | 0.192*** | 0.056 |
| VirusSearch | cov | Finance | −0.947*** | 0.031 |
| VirusSearch | cov | SocDist | −0.566*** | 0.051 |
| HealthStatus | cov | Mobility | −0.278*** | 0.047 |
| HealthStatus | cov | Finance | −0.269*** | 0.062 |
| HealthStatus | cov | SocDist | −0.928*** | 0.010 |
| Mobility | cov | Finance | −0.676*** | 0.034 |
| Finance | cov | SocDist | 0.450*** | 0.059 |
| PsySearch | cov | jAdultContent | 0.247*** | 0.047 |

*Note:* *p<0.1; **p<0.05; ***p<0.01

**Table 11:** The fitting results of the SEM model for the Japanese data.

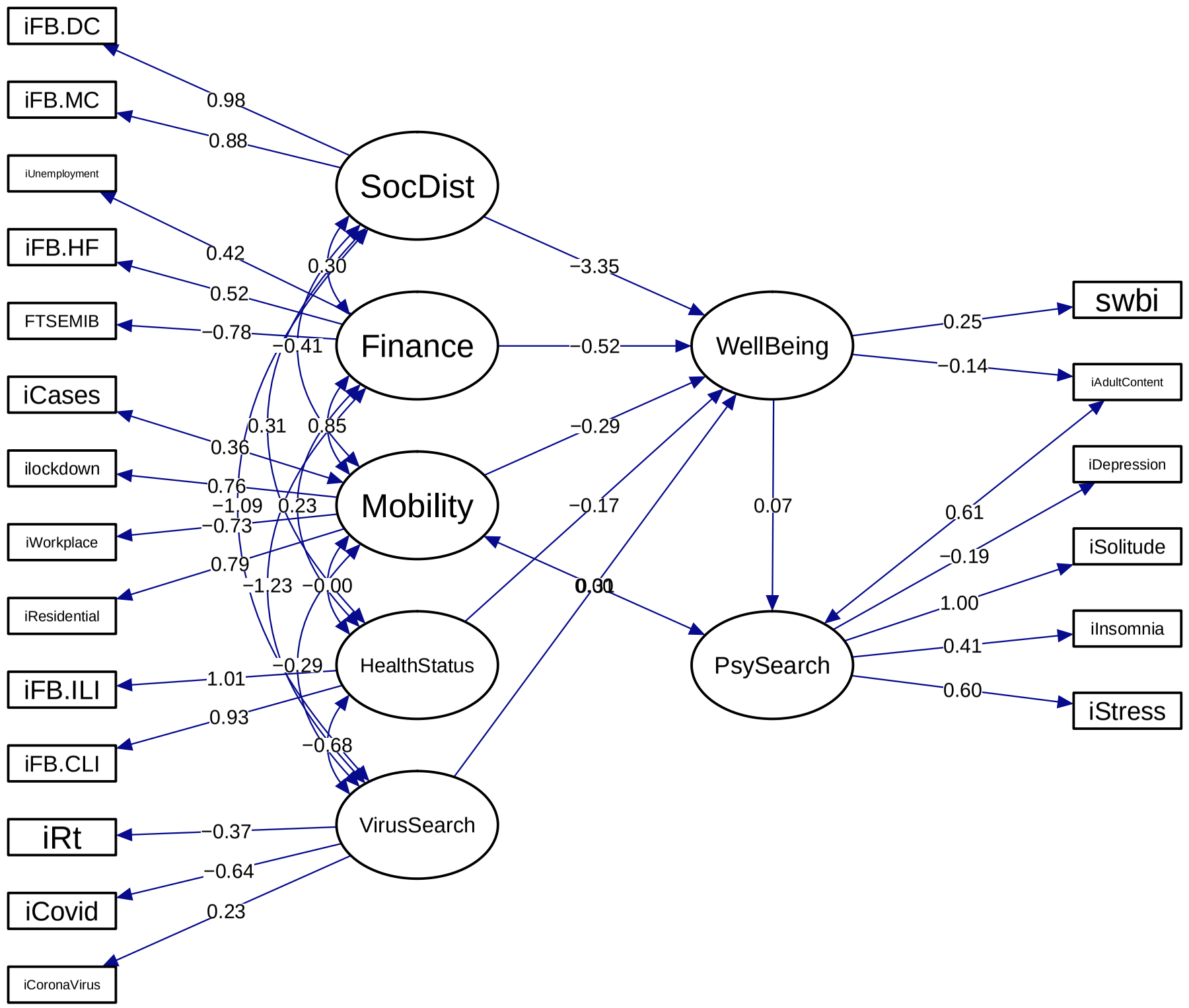

**Figure 10:** Output of the SEM fitting for SWB-I.

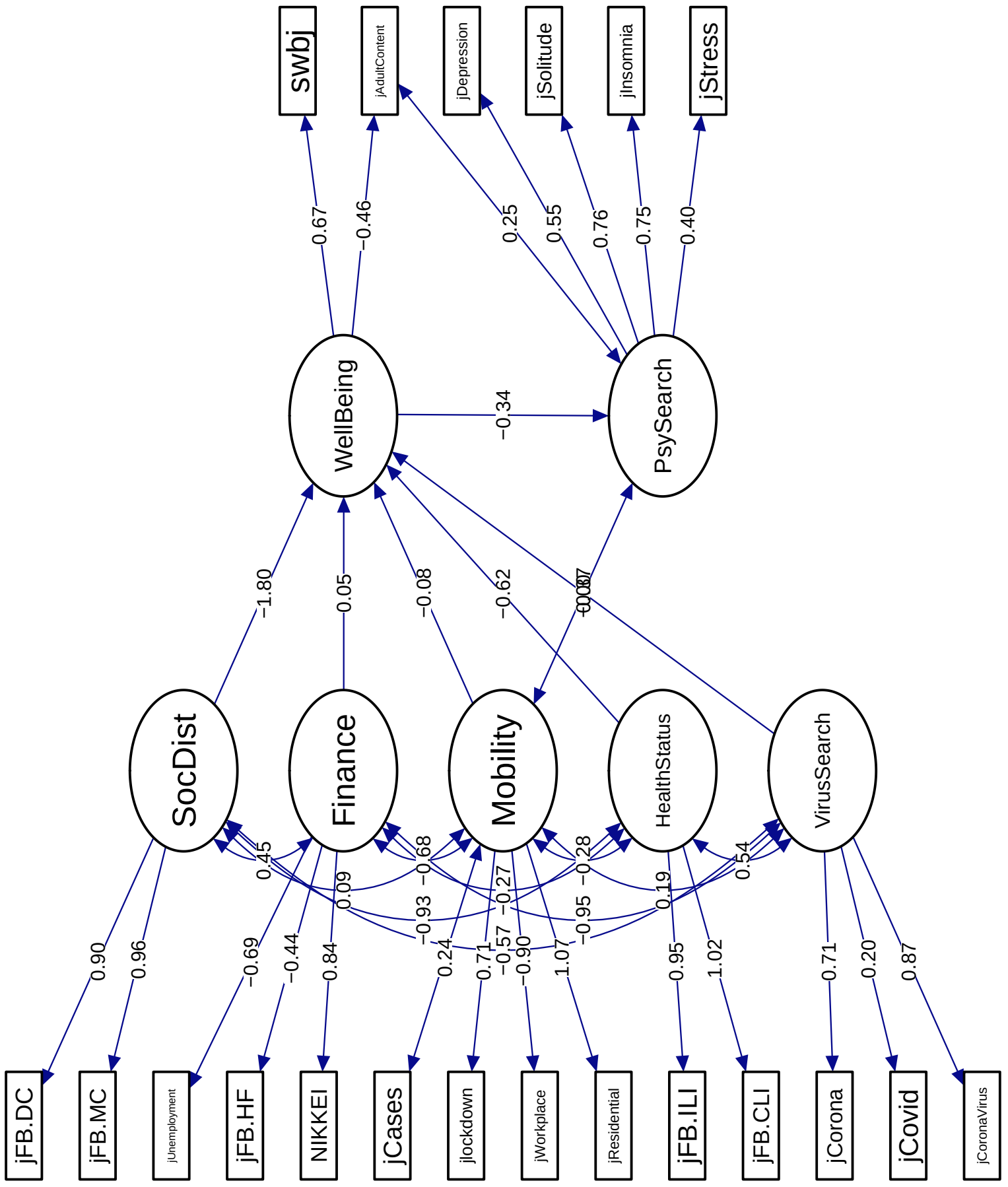

**Figure 11:** Output of the SEM fitting for SWB-J.

# 9 Limits of This Study

This study was limited in its extent mainly due to data availability and it only applies to two countries: Italy and Japan. It would be nice to extend the same kind of analysis to, at least, USA and a few other European and Latin American countries that experienced quite dramatic impact of the pandemic in terms of lost lives and show remarkable economic and cultural differences . We hope to fill this gap in the near future. The supervised sentiment analysis approach is accurate but requires human coding in the native language of the tweets, which is not a scalable task without resources.

Though we managed to keep the set of potential explanatory variables similar between the countries, this may not be necessarily as easy in a more global application of the analysis. For example, in Russia the platform VK replaces Facebook, but no similar survey data are available. China, on the other hand, is completely excluded because of the different structure of social network and data accessibility (i.e., Twitter is replaced by Sina Weibo, Google is simply not available, as well as Facebook), Still it is not impossible to think of elaborating equivalent subjective well-being indexes.

Finally, the causal relationships between the explanatory variables and the Twitter subjective well-being indexes have been only partially explored in this work throughout the structural equation modeling, and probably quite a few unobserved (but non necessarily unobservable) variables exist that could improve the analysis overall.

Despite the above limitations, we think that this study has shown several meaningful correlative (and possibly causal) effects and shed some light on inter-dependencies and specificities that may help in understanding and interpreting the different reactions of countries to apparently similar shocks.

# Acknowledgments

The data collection has been performed within the Japan Science and Technology Agency CREST (Core Research for Evolutional Science and Technology) project, grant n. JPMJCR14D7. This work was performed also in collaboration with the Waseda Institute of Social Media Data (WISDOM), especially for the part concerning the training set human coding.

# Competing and/or Conflict of Interests

SI is currently on leave from the University of Milan and employed at the Joint Research Centre of the European Commission. SI conducted the statistical analyses while at the University of Milan and wrote the paper while at the Joint Research Centre.

# Author Contributions

SI and GP conceived the project. TC, AH, SI and GP designed the study. TC and AH wrote the intercultural and socio-linguistic analysis. SI and GP did the statistical analyses of the data. TC, AH, SI, and GP interpreted the results and wrote and revised the article.

# Data Availability

All data and R scripts are available here https://github.com/siacus/swbCovid.